\documentclass{aa}
\usepackage[dvips]{graphicx}
\newcommand{\nm}{nm}
\newcommand{\kms}{{\,km\,s$^{-1}$ }}

\newcommand{\bi}{\begin{itemize}}
\newcommand{\ei}{\end{itemize}}
\newcommand{\Lstar}{\mbox{$L_* $}}

\newcommand{\Lsun}{\mbox{$L_{\odot} $}}
\newcommand{\Msun}{\mbox{$M_{\odot} $}}

\newcommand{\Teff}{\mbox{$T_{\rm eff} $}}
\newcommand{\La}{\mbox{L$\alpha $}}
\newcommand{\Ha}{\mbox{H$\alpha $}}
\newcommand{\Hb}{\mbox{H$\beta $}}
\newcommand{\Ib}{\mbox{I$_{\beta} $}}
\newcommand{\Fb}{\mbox{F$_{\beta} $}}

\newcommand{\Ne}{\mbox{N$_e $}}
\newcommand{\Nh}{\mbox{N$_H $}}

\newcommand{\he}{He$\,$2-436}
\newcommand{\wray}{Wray$\,$16-423}

\def \eg{{\it e.g.}}
\def \cmc{cm$^{-3}$}
\def \hi{H$\,\textsc{i}$}
\def \hei{He$\,\textsc{i}$}
\def \heii{He$\,\textsc{ii}$}
\def \cii{C$\,\textsc{ii}$}
\def \ciii{C$\,\textsc{iii}$]}
\def \ciiip{C$\,\textsc{iii}$}
\def \civ{C$\,\textsc{iv}$}
\def \nii{[N$\,\textsc{ii}$]}
\def \niii{N$\,\textsc{iii}$}
\def \oi{[O$\,\textsc{i}$]}
\def \oii{[O$\,\textsc{ii}$]}
\def \oiip{O$\,\textsc{ii}$}
\def \oiii{[O$\,\textsc{iii}$]}

\def \neiii{[Ne$\,\textsc{iii}$]}

\def \sii{[S$\,\textsc{ii}$]}
\def \siii{[S$\,\textsc{iii}$]}
\def \mgi{Mg$\,\textsc{i}$]}
\def \cliii{[Cl$\,\textsc{iii}$]}
\def \ariii{[Ar$\,\textsc{iii}$]}
\def \ariv{[Ar$\,\textsc{iv}$]}
\def \kiv{[K$\,\textsc{iv}$]}
\def \la{$\lambda$}
\begin{document}

   \thesaurus{08             
             (08.16.4        
              09.16.1        
              09.16.2 He~2-436     
              09.16.2 Wray~16-423  
              11.04.2        
              11.09.1 Sagittarius)  
}

\title{A radio-continuum and photoionization-model study of the two planetary nebulae in the Sagittarius dwarf galaxy}
\titlerunning{Planetary nebulae in Sagittarius dwarf galaxy}
\author{G. Dudziak\inst{1}
\and
D. P\'{e}quignot\inst{2}
\and
A. A. Zijlstra\inst{3}
\and
J. R. Walsh\inst{4}
}
\authorrunning{Dudziak et al.}
\offprints{D. P\'equignot}
\institute{
Department of Physics and Applied Physics, University of Strathclyde, 
Scotland.
E-mail: gregory.dudziak@strath.ac.uk 
\and
Laboratoire d'Astrophysique Extragalactique et de Cosmologie associ\'{e}
au CNRS (UMR 8631) et \`{a} l'Universit\'{e} Paris 7,
DAEC, Observatoire de Paris-Meudon, 
F-92195 Meudon Principal Cedex, France.
Email: daniel.pequignot@obspm.fr
\and
Department of Physics,
University of Manchester Institute of Science and Technology,
P.O. Box 88, 
Manchester, M60 1QD, 
United Kingdom.
Email: aaz@iapetus.phy.umist.ac.uk
\and
Space Telescope European Co-ordinating Facility,
European Southern Observatory,
Karl-Schwarzschild Strasse 2, 
D-85748 Garching bei M\"{u}nchen,
Germany.
E-mail: jwalsh@eso.org
}

\date{ }
\maketitle

\begin{abstract}
Radio continuum observations at 1.4, 4.8 and 8.6~GHz of the two 
Planetary Nebulae (PNe) in the Sagittarius dwarf galaxy reveal the 
elongated shape of Wray~16-423 and the extreme compactness of He~2-436. 
It is confirmed that He~2-436 is subject to local dust extinction. 

Photoionization models for both PNe are obtained from two different 
codes, allowing theoretical uncertainties to be assessed. Wray~16-423, 
excited by a star of effective temperature 1.07$\times10^5$K, 
is an ellipsoidal, matter-bounded nebula, except for a denser sector 
of solid angle 15$\%$. He~2-436, excited by a 7$\times10^4$K star, 
includes two radiation-bounded shells, with the very dense, low-mass, 
incomplete, inner shell possibly corresponding to a transitory event. 
The continuum jump at the He$^+$ limit ($\lambda$22.8nm) agrees with NLTE 
model stellar atmospheres, despite the Wolf-Rayet nature of the stars. 
Both stars are on the same (H-burning) evolutionary track of 
initial mass (1.2$\pm$0.1)~\Msun\ and may be twins, with the PN ejection 
of Wray~16-423 having occured $\sim$~1500 years before He~2-436.

The PN abundances re-inforce the common origin of the 
parent stars, indicating almost identical depletions with respect 
to solar for O, Ne, Mg, S, Cl, Ar, and K (-0.55$\pm$0.07~dex), 
and strong overabundances for carbon, particularly in He~2-436. 
\hei\ lines consistently point to large identical overabundances for helium 
in both PNe. An excess nitrogen makes Wray~16-423 nearly a Type~I~PN. 

These PNe provide a means to calibrate both metallicity and age of 
the stellar population of Sagittarius. They confirm that the youngest, 
most metal-rich population has an age of 5Gyr and a metallicity of 
[Fe/H]$=-0.55$, in agreement with the slope of the red giant branch. 

\end{abstract}

\section{Introduction}
Zijlstra \& Walsh (\cite{zijwal96}) discovered 
that two previously catalogued Planetary Nebulae (PNe) belonged to the 
newly recognized Sagittarius dwarf galaxy (Ibata et al. \cite{ibagil94}, 
\cite{ibagil95}). In the course of a radial velocity survey 
(Zijlstra et al. \cite{zijack97}), He~2-436 and Wray~16-423 
were found to coincide in both position and velocity 
with this galaxy. Optical spectroscopy for both objects 
(Walsh et al. \cite{waldud97}, hereafter Paper~I), showed 
a $\sim$ --0.5dex depletion of N, O, Ne with respect to 
Galactic PNe (Kingsburgh \& Barlow \cite{kinbar94}), 
confirming that they were members of Sagittarius. 
Wolf-Rayet ([WR]) features were detected in both spectra. 

Given that these objects are the closest extragalactic 
PNe ever identified [distance to Sagittarius is taken as 25kpc 
(Ibata et al. \cite{ibagil95}), although Mateo (\cite{mat98}) gives 24kpc]  
and belong to a stellar system 
with a star formation history different from that of the Milky Way, 
it was decided to obtain radio continuum data and to improve on 
the abundances and other physical quantities of the nebulae and 
central stars by means of photoionization models. 

\section{Radio observations}

\subsection{Wray~16-423}

Wray~16-423 was observed with the Australian Telescope Compact Array 
on 1997, June 19th, simultaneously at 
8.6 and 4.8~GHz over a 12-hour integration. The array was in its
`6A' configuration with baselines ranging from 330m to 5940m,
maximizing the angular resolution. The radio source 1933$-$400 was 
used for phase calibration. The amplitude calibration was based on 
1934$-$638, adopting the new fluxes of 5.826~Jy and 2.814~Jy 
at 4.8~GHz and 8.64~GHz respectively. (The new scale differs 
from the previous one, changing the spectral index
between 3 and 6~cm by 0.3. This scale agrees with
the one used in the Northern hemisphere to within 1--2\%.)

The data were analyzed using the Miriad and AIPS packages. 
A full calibration of instrumental polarization was done
to optimize the flux calibration.  Tests done
during the observations showed a pointing accuracy of better
than $10\arcsec$.  Altitude-dependent pointing errors were 
minimized as the phase calibration was close to the target source.
The calibrated data
were Fourier transformed to produce images and cleaned to replace the
'dirty' point spread function by a Gaussian beam. 
A signal-to-noise ratio of 300 was measured before self-calibration, 
confirming the quality of the data.

The FWHM of the resulting Gaussian beam was $3.0\arcsec \times
1.8\arcsec$ at 6~cm and $1.7\arcsec \times 1.0\arcsec$ at 3~cm, with
position angle near zero degrees, owing to the 12-hour
integration. Because the point-spread function is mathematically
known, sizes can be obtained even for sources several times smaller 
than the beam for sufficiently high signal-to-noise. 
Wray~16-423 was found to be slightly 
resolved at 3~cm, with equivalent FWHM of a Gaussian source of
$0.6\arcsec \times 0.3\arcsec$ at position angle 58 degrees
(measured from North to East on the sky). At 6~cm a size of
$0.85\arcsec \times 0.3\arcsec$ was found with the same position
angle. The PN appears unresolved along its short axis but the observed 
elongation is not an instrumental artefact, given that 
the position angle is quite different from that of the beam. 
Thus, Wray~16-423 may have an elliptical or otherwise
elongated morphology with axial ratio of order 2:1 or more. 
Correction factors between FWHM Gaussian diameters and
uniform disk diameters for slightly resolved sources are provided by 
Bedding \& Zijlstra (\cite{bedzij94}) 
and van Hoof (\cite{hoof00}): for Wray~16-423 the Gaussian 
diameter should be multiplied by a factor of 1.6, giving approximately
$1\arcsec$ for the major axis. In Paper~I,  
a ground-based \oiii\ image of Wray~16-423 was deconvolved and a 
1.2$\times$0.8$''$ ellipse with position angle 50$^\circ$ was found.
The radio and optical descriptions are therefore in agreement within 
uncertainties. 

The total flux of Wray~16-423 was 4.72~mJy at 6~cm and 4.31~mJy at
3~cm.  The flux accuracy, limited by the noise in the map
(0.06~mJy/beam) and the systematic uncertainty in the flux
scale, is 0.20~mJy at the 3-$\sigma$ level. The
spectral index is found to be $-0.08$, very close to the theoretical
value of $-0.1$ for optically thin radio emission, thus confirming the
accuracy of the relative flux calibration.

\subsection{He~2-436}

He~2-436 was observed at the VLA on October 1, 1998, using
15-minute snapshots at 3.6~cm (8.4~GHz), 6~cm (4.89~GHz) and 20~cm
(1.46~GHz). The array was in the 'B' configuration. 3C286 was used
as flux calibrator and 1924-292 the phase calibrator. Due to the
short integrations, the calibration is less accurate than for Wray~16-423. 

After cleaning and self-calibration, He~2-436 was detected at all
bands although very faint at 20~cm. The flux was determined as $0.6
\pm 0.2$ mJy at 20~cm, $3.9 \pm 0.2$ mJy at 6~cm and $4.9 \pm 0.2$ mJy
at 3.6~cm. The uncertainty comes from the noise in the map (around 0.1 
mJy/beam but 0.16 mJy/beam at 20~cm) and an estimate of the uncertainty
in the flux calibration.  The source was unresolved. At 3.6~cm where
the beam was $2\arcsec \times 0.8\arcsec$ (elongated NS), an upper
limit to the size of $0.2 \arcsec$ in the EW direction was found.

The turn-over frequency between optically thick and optically thin 
emission may occur near 8~GHz, suggestive of a high density PN. 
Assuming $T_b = T_e$ at 20~cm, the observed flux of 0.62 mJy 
corresponds to a source
diameter of $0.12 \arcsec$. The diameter may be larger if the optical
depth is not constant across the nebula. The spectral index between 6
and 20~cm is $-0.7$, compatible with an $r^{-2}$
density distribution. The interpretation may not be unique but 
a density gradient is detected in He2-436. 

The radio position of He~2-436 is RA $19^h 32^m 06.72^s$, DEC
$-34^{\circ}12^{\prime}57.3^{\prime\prime}$ (J2000), in agreement 
with the optical position given by Walsh et al. (\cite{waldud97}). 
This differs from the J2000 listed position in the 
Strasbourg-ESO catalogue of Galactic PNe 
(Acker et al. \cite{ackoch92}) but agrees with the star position 
in the finding chart. 

\section{Photoionization models}

\subsection{Computations}

\subsubsection{The codes}

Models for these PNe were obtained from two photoionization 
codes which differed in their numerical methods and approximations: 
on the one hand, the code developed by 
J. P. Harrington (Harrington et al. \cite{harsea82}), 
as further modified by R. E. S. Clegg (Clegg et al. \cite{clehar87}) 
-- hereafter the HC code -- and updated for some atomic data 
according to the package developed by Shaw \& Dufour (\cite{shaduf95}); 
on the other hand, the NEBU code, a descendent of the code of 
G. Stasinska, S. M. Viegas-Aldrovandi and D. P\'equignot, as further 
developed by P\'equignot (Petitjean et al. \cite{petboi90}; 
Morisset \& P\'equignot \cite{morpeq96}). An overview of the characteristics 
of these and other codes is presented in Ferland et al. (\cite{ferlan95}). 

Atomic data, notably from the IRON project, are reviewed by 
Storey (\cite{storey97}). More recent references of interest are 
Ramsbottom et al. (\cite{rambel97}, \cite{rambel99}) for \ariv\ 
and \cliii. The total recombination coefficients have now been re-assessed 
for C, N and O (Nahar \& Pradhan \cite{nahpra97}; Nahar \cite{nahar99}), 
but are not yet known for some 
of the S, Cl, and Ar ions (\eg, Nahar \cite{nahar95}). In 
NEBU, a few coefficients were empirically calibrated 
from a new unpublished model of the PN NGC~7027: the 
S$^{2+}$ $\rightarrow$ S$^+$ and Ar$^{3+}$ $\rightarrow$ Ar$^{2+}$ total 
recombination coefficients were increased by factors 1.7 and 5 
respectively. Note that, for S$^{3+}$ $\rightarrow$ S$^{2+}$, 
Nahar (\cite{nahar95}) obtained a factor 5 increase.

\subsubsection{Primary continua}

In the HC computation, the central star energy distribution was 
taken from the NLTE model atmospheres of Clegg \& Middlemass 
(\cite{clemid87}). Given the luminosity \Lstar\ and the effective 
temperature \Teff, the surface gravity $g$ was bound to give 
a stellar mass in the range 0.55 -- 0.65~\Msun, 
encompassing 80\% of the PN nuclei 
(Stasinska et al. \cite{stagor97}). 

However concerns may be expressed about standard 
stellar atmosphere models, particularly when applied to [WR] stars 
(\eg, Kudritzki \& M\'endez \cite{kudmen93}). $m_{\rm V}$
In the NEBU computation, the primary radiation was given a 
black-body shape (defined by $L_B$, $T_B$, equivalent to 
\Lstar, \Teff), except that the flux shortward of \la22.8\nm\ 
could be multiplied by an arbitrary factor $f_4$. This 
freedom allowed some consequences of varying the hardness 
of the ionizing continuum to be studied. Longward of \la22.8 
the black-body flux matches the 
model-atmosphere flux of a star with \Teff\ = $T_B$ -- 1.5$\times10^4$K 
and mass 0.6~\Msun, appropriate for a $M$ = 1.2~\Msun\ parent star 
with $Z$ = 0.004 (Sect.~4.6). This relation will be used 
to convert $T_B$'s into \Teff's.

\subsubsection{Gas distributions}

Both computations were done in spherical symmetry, with filling 
factor unity and uniform elemental abundances. Departure from 
spherical symmetry was approximately taken into account 
by combining two spherical cases (two-sector models). 
In the HC computation, the density was constant, sometimes constant 
by steps (with linear transitions), throughout the nebula. 
In the NEBU computation, the gas pressure was 
constant, particularly in the vicinity of the ionization fronts, or 
constant by steps with analytically smooth transitions. 
No attempt was made to model the effect of the dust possibly mixed 
with the ionized gas. 

\subsection{Observations}

\subsubsection{Optical data}

A considerable asset of remote PNe from the standpoint of 
photoionization modeling is that their small apparent size 
allows global spectra to be secured. 
Optical line fluxes were taken from 
Paper~I. In col.~6 of Table~2 and col.~4 of Table~4 below, 
the 1-$\sigma$ errors are from the Gaussian fits to the lines 
and do not include possible systematic errors. In Wray~16-423, 
the \Hb\ flux was corrected for \heii. 

Concerning blends, \ariv\ 471.3\nm\ was corrected for \hei\ 471.1\nm\ 
($\sim$ 34\% and 85\% of the blend in Wray~16-423 and He~2-436 
respectively, see below and Sect.~4.3), \sii\ 406.9\nm\ for \ciiip+\oiip\ 
(2.87$\times$I$_{\lambda418.7}$+0.518$\times$I$_{\lambda465.1}$, 
typically 23\% and 15\% in Wray~16-423 and He~2-436 resp.), 
and \sii\ 407.6 for \cii+\oiip\ 
(0.074$\times$I$_{\lambda426.7}$+0.349$\times$I$_{\lambda465.1}$, 
typically 18\% and 14\% resp.) 
(see P\'equignot et al. \cite{peqpet91}; Storey \cite{storey94}; 
Liu et al. \cite{liusto95}; Davey et al. \cite{davsto00}). 
These corrections were applied consistently for each model. Thus 
the computed \cii~426.7\nm\ flux, rather than the observed one, was used, 
considering the large observational uncertainty. 
(The fluxes of \oii\ 732.0\nm\ and 733.0\nm\ should 
be exchanged in Table~4 and Table~5 of Paper~I; only the sum of 
this multiplet is used for diagnostics. In Table~4 of Paper~I, 
the $\lambda$471.1 flux should read 1.40, not 1.49). 

In Wray~16-423, after correcting for \siii~372.2\nm, 
the flux quoted for \hi~372.2\nm\ (B14) is still significantly 
larger than the theoretical one, whereas all other Balmer lines from 
B9 to B19 agree with theory (Storey \& Hummer  \cite{stohum95})  
within uncertainties (B16 is too strong, but blended  
with \hei, Sect.~4.3). No other blend with B14 is 
known but the line is close to \oii~372.6\nm\ and the excess flux 
of B14 was attributed to \oii\ (with enhanced uncertainty). 
A similar excess, present in He~2-436, was treated in the same manner, 
but the correction to \oii\ is tentative owing to the 
poorer definition of the continuum level. In these PNe, 
the \oii\ doublet ratio is little sensitive to 
physical conditions. These corrections to $\lambda372.6$ 
lead to new observed ratios in better agreement with theory. 

The asymmetrical feature \la723.6 
(noted \ariv\ in Paper~I) is dominated 
by \cii(3) 723.1+3.6+3.7\nm, whose strength is 
1.04 times that of \la426.7 in Case~B and small in 
Case~A (Davey et al. \cite{davsto00}). Moderate and 
``moderately small'' departures from Case~B are expected for 
Wray~16-423 and He~2-436 respectively. Indeed, after removing 
\ariv\ 723.7\nm\ (below), the ratio 
I$_{\lambda723.6}$/I$_{\lambda426.7}$ is 0.58 and $\sim$0.82 in Wray~16-423 
and He~2-436 respectively. 
Removing \cii~658.0\nm\ from \nii~658.3\nm, assuming 
half Case~B, leads to an improved \nii-doublet ratio. 

For densities of interest, the sum of the red 
\ariv\ multiplet is 3.3 times the intensity of \ariv\ 726.3\nm\ 
(Mendoza \& Zeippen \cite{menzei82}), detected in Wray~16-423 at low 
signal to noise. In He~2-436, \ariv\ 726.3\nm\ and 717.1\nm, 
of similar intensities, were not detected: 
a 3-$\sigma$ upper limit to the multiplet 
was taken as 3 times the 2-$\sigma$ limits. 

The observed \ariii\ ratio I$_{\lambda713.6}$/I$_{\lambda775.1}$ is 
$\sim$~5.3 in both objects (Paper~I) and the theoretical value is 4.15 
(Mendoza \cite{mendoz83}). 
The adopted sum for the multiplet was based on \la713.6, using 
a doublet ratio intermediate between the observed and theoretical 
values and increasing the error bar.  
In He~2-436, the ``2-$\sigma$ detections'' of the 
\cliii\ lines (Paper~I) were replaced by 3-$\sigma$ upper limits.

\subsubsection{Dust extinction and absolute \Hb\ fluxes}

The radio fluxes for Wray~16-423 lead to 
log~$\Ib\ = -11.89 \pm 0.02$ 
and the observed log~F$(\Hb) = -12.09 \pm 0.03$ implies an extinction 
c$_{\rm radio}$ = 0.20$\pm$0.04, which compares very well with the 
extinction derived from the Balmer decrement, 
c$_{\rm Balmer}$ = 0.20$\pm$0.04 (Paper~I). 
Considering this excellent agreement and the accuracy of the 
radio fluxes, the uncertainty inherent to the reddening 
correction should be negligible and the de-reddened absolute 
\Hb\ flux be determined to better than 0.03dex. 

The shape of the radio continuum of He~2-436 indicates that the fully 
optically thin regime may occur above 8~GHz, the highest 
frequency available. A simple extrapolation suggests 
an optically thin flux $\approx$~5.4 mJy, which, in combination 
with F$(\Hb)$, would lead to c$_{\rm radio} = 0.50 \pm 0.07$, 
smaller than c$_{\rm Balmer} = 0.61 \pm 0.05$. 
The interpretation of the Balmer decrement may be 
affected by extinction internal to the H$^+$ region 
and, particularly, by local uneven absorption due to 
a neutral envelope, as in the case of NGC~7027 
(\eg, Robberto et al. \cite{robcla93}) 
since the ionized mass of He~2-436 is small. 
Now, in case of uneven dust extinction, the Balmer-decrement method 
{\sl underestimates} the true average extinction, 
at variance with what observational data suggest. 
Thus, until radio fluxes are obtained at higher frequencies, the 
Balmer-decrement reddening is adopted on account of the excellent agreement 
between c$_{\rm radio}$ and c$_{\rm Balmer}$ in the case of Wray~16-423. 

In fact it is anticipated that the model results (Sect.~3.4, Table~4) will 
indicate that, assuming c $= 0.61$, the de-reddened \Ha/\Hb\ of Paper~I is 
off by 2\% for He~2-436, hence a new best reddening correction with reduced 
error bar c$_{\rm Balmer} = 0.58 \pm 0.03$, still compatible with the value 
of Paper~I but somewhat closer to c$_{\rm radio}$. In the following, 
this new c$_{\rm Balmer}$ will be used to obtain de-reddened optical fluxes 
relative to \Hb. The present high-frequency radio fluxes are not too 
much affected by self-absorption and are more accurate 
than the \Hb\ flux, taken in this case from 
Webster (\cite{webste83}): most of the uncertainty on 
c$_{\rm radio}$ and thus presumably most of the difference between 
c$_{\rm radio}$ and the new c$_{\rm Balmer}$ arise from the uncertainty 
of 15\% on \Hb, not from the extrapolation of the radio fluxes 
to optically thin frequencies. It can therefore be suspected 
-- as models will confirm -- that the best 
estimate for the \Hb\ flux corresponds to the lower end 
of the interval allowed by the error bars, namely Webster's flux 
divided by 1.15. 

\subsection{Wray~16-423}

\subsubsection{HC computation}

Wray~16-423 is still a rather dense and young PN and 
it is relatively safe to assume that nebular material is 
present along any direction from the star. The total covering 
factors will be taken equal to unity. 

The one-component constant-density model is described 
in col.~2 of Table~1 
and the line intensities are given in col. 2 of Table~2 (``Thin'').
The outer radius of this model roughly corresponds to the 
observed mean apparent diameter of 1.0\arcsec. 
This ``thin'' model (optical depth $\tau_{13.6}$ = 7 at 13.6~eV) 
is a poor fit to low-ionization line fluxes. 

\begin{table}
\caption[]{Models for Wray~16-423}
\begin{flushleft}
\begin{tabular}{lcc}
Model                         &  HC   &   NEBU   \\
Distance/kpc                  &  25   &    25    \\ 
(\Teff\ or $T_B$)/10$^3$K     &  90   &   125    \\
(\Lstar\ or $L_B$)/10$^3$\Lsun & 4.60  &  5.04$^a$ \\  
log~$g$ \ \ \ \ \ \ \ \ ($f_4$) &  5.3  &  (0.31)  \\
N$_{13.6}$/10$^{47}$s$^{-1}$  & 3.11  &  2.95    \\
Filling factor      &    1.0        &     1.0   \\
Mass/10$^{-2}$\Msun &  18.8         &    24.8   \\
\multicolumn{2}{l}{Radius/10$^{-3}$pc:}       &     \\
(Thin sector)       &    33 - 62    &   23 - 80     \\
(Thick sector)      &    33 - 73    &   23 - 51     \\
\multicolumn{3}{l}{\Nh/10$^4$\cmc\ \ (and covering factor):}  \\
(Thin)              & 0.60 (0.90)   & 0.36 (0.83) \\
(Thick)             & 0.60 (0.10)   & 0.95 (0.17) \\
\multicolumn{3}{l}{Gas pressure/10$^{-8}$CGS:}  \\
(Thin)              &       -   &  1.4    \\
(Thick)             &       -   &  3.5    \\
$\tau_{13.6}$(Thin) &     7.0   &  6.7    \\
\multicolumn{2}{l}{Abundances by number:}  &   \\
H                                   & 1.00 & 1.00 \\
He                                  & 0.107& 0.107$^b$ \\
C \ \ ($\times$10$^{5}$)            & 67.6 & 73.7 \\
N \ \ ($\times$10$^{5}$)            & 5.79 & 4.60 \\
O \ \ ($\times$10$^{5}$)            & 20.4 & 21.4 \\
Ne \ ($\times$10$^{5}$)             & 3.33 & 3.59 \\
Mg ($\times$10$^{5}$)               &  -   & 0.95 \\
S \ \ \ ($\times$10$^{5}$)          & 0.61 & 0.440 \\
Cl \ \ ($\times$10$^{5}$)           &  -   & 0.0077 \\
Ar \ ($\times$10$^{5}$)             &  -   & 0.090 \\
K \ \ ($\times$10$^{5}$)            &  -   & 0.0045 \\
\end{tabular}

\ \ $^a$ 4.22, correcting for $f_4$ (see Sect.~3.3.3).\\
\ \ $^b$ Adopted He abundance will be 0.108 (Sect.~4.3). \\
\end{flushleft}
\end{table}

Consideration was given to an optically thick component obtained by 
increasing the outer radius from 0.062 to 0.073~pc. 
On its own, this ``thick'' model (col.~3 of Table~2) 
does not reproduce the observations.  
Wray~16-423 cannot be spherically symmetric: if the optical depth 
is large enough to account for \oi, then \oii\ is too large. 

A composite of the thin and thick cases (Table~2, col.~4), 
accounting for the \oii\ lines,  
is an improvement in that not only \oii, but \sii\ and \oi\ as well 
are now better matched, although \oi\ is still underestimated. 
\cii\ 426.7\nm\ appears 
underestimated by 30$\%$, still within the 1-$\sigma$ error bar: 
this result should be considered a success of this model 
since carbon is determined by the global energy budget, whereas 
\Teff\ is constrained by the \oiii\ ratio. 
The \sii\ line ratios are now not as closely matched as in the previous 
``thin case'', suggesting that low-ionization lines are 
not produced at a unique density.

\subsubsection{NEBU computation}

The low-ionization line fluxes were improved by introducing a thick sector 
in the HC model. Nonetheless, assuming a constant density throughout the 
nebula, the radiation-bounded sector was left as the long axis. 
In elongated Galactic PNe, the short axis generally 
corresponds to the dense, optically thicker ``equator''. 

\begin{table*}
\caption[]{Observations and model predictions for Wray~16-423}
\begin{flushleft}
\begin{tabular}{lrrr|rr|rrrr}
      & \multicolumn{3}{c}{HC} \vline &
\multicolumn{2}{c}{Observation} \vline & \multicolumn{4}{c}{NEBU} \\
\cline{2-10}
                 & Thin  & Thick &  Model  & Flux &  1$\sigma$ Err &  Thin &  Thick & Model& Mod/Obs \\
Covering factors:    & 0.90 & 0.10 & 1.00  & &                    &   0.835 & 0.165 &  1.00 &        \\ 
\multicolumn{3}{l}{\underline{Absolute fluxes}}  &&& && & &\\
log~\Ib(erg cm$^{-2}$s$^{-1}$)
                       & -11.89 & -11.72  & -11.87 & -11.89 &0.03& -11.93 & -11.74 & -11.89 & 1.00$\pm$.07\\
4.8 GHz (mJy)            &   4.5 &   6.7  &   4.8 &   4.72 & 0.10  &   4.35 &   6.37  &   4.69  & 0.99$\pm$.02\\
8.6 GHz (mJy)            &   4.3 &   6.4  &   4.5 &   4.31 & 0.10 &   4.12 &   6.12  &   4.45 & 1.03$\pm$.02\\
\multicolumn{3}{l}{\underline{Relative line fluxes} (wavelengths in \nm)}  &&&&&& \\
 \hi\ 486.1             & 100.00 &  100.00 & 100.00 & 100.00 &  -  & 100.00 &100.00 &  100.00 &  -         \\
 Cont. 364.2 (/\nm)     &    -   &     -   &     -   &  48.8 &3.00 &   51.0 &  48.9 &   50.5 & 1.03$\pm$.06\\
 Cont. 364.8 (/\nm)     &    -   &     -   &     -   &  12.0 &1.00 &   12.8 &  11.4 &   12.4 & 1.03$\pm$.08\\
 \hi\ 656.3             & 284.00 &  300.00 & 287.00 & 284.00 &1.60 & 285.00 &291.00 & 286.00 & 1.01$\pm$.01\\
 \hei\ 447.1            &   5.51 &    5.13 &   5.46 &  5.44  &0.23 &   5.32 &  5.89 &   5.45 & 1.00$\pm$.04\\
 \hei\ 587.6            &  15.40 &   13.90 &  15.20 & 16.40  &0.10 & 15.80  & 18.10 & 16.30  & 1.00$\pm$.01\\
 \heii\ 468.6           &  10.70 &    7.37 &  10.30 &  11.40 &0.14 & 12.40  &  8.10 & 11.40  & 1.00$\pm$.01\\
 \cii\ 426.7            &   0.35 &    0.37 &   0.35 &   0.49 &0.15 &   0.47 &  0.51 &  0.48  &  .98$\pm$.30\\
 \ciiip\ 418.7          &   -    &    -     &   -     & $<$0.22 & 3$\sigma$ &  0.14 & 0.11 & 0.13 &$>$ 0.59\\
 \ciii\ 190.9+0.7        & 739.00 &   879.00 &  759.00 &  -    &   - & 749.00 & 906.00 & 783.00 &    -      \\
 \civ\ 154.9            & 714.00 &   516.00 &  687.00 &  -    &   - & 630.00 & 467.00 & 594.00 &    -      \\
[0.2cm]
 {\nii} 658.4+4.8   &  11.40 &  103.00 &  24.30 &  22.70 & 0.16  &   8.31 &   70.80 & 22.90  & 1.01$\pm$.01\\
 {\nii} 575.5      &   0.25 &    2.12 &   0.51 &   0.53 & 0.05  &   0.16 &    1.51 &  0.48  &  .90$\pm$.10\\
 {\oi} 630.0+6.3    &   0.01 &    5.05 &   0.72 &   2.95 & 0.10  &   0.02 &    7.05 &   1.67 &  .56$\pm$.02\\
 {\oii} 372.6      &  11.20 &   77.50 &  20.40 &  20.30 & 0.40  &  12.50 &   47.70 &  20.70 & 1.02$\pm$.02\\
 {\oii} 372.9      &   4.57 &   32.60 &   8.57 &   8.84 & 0.27  &   5.85 &   17.80 &   8.65 &  .98$\pm$.03\\
 {\oii} 732.0+3.0   &   2.42 &   16.20 &   4.36 &   4.68 & 0.11  &   2.10 &   13.40 &   4.74 & 1.01$\pm$.02\\
 \oiip\ 465.1+     &   -    &    -    &    -   & $<$0.56 & -    &   0.25 &    0.23 &   0.24 &   $>$ 0.43  \\
 {\oiii} 500.7+495.9 &1450.00 & 1340.00 &1438.00 &1460.00 & 6.00 &1474.00 & 1415.00 &1460.00 & 1.00$\pm$.00\\
 {\oiii} 436.3      &  14.00 &   13.50 &  13.90 &  13.70 & 0.22 &  13.60 &   14.00 &  13.70 & 1.00$\pm$.02\\
 {\neiii} 386.8+396.7&  94.60 &  101.40 &  95.60 &  97.70 & 0.90 &  96.70 &   99.90 &  97.70 & 1.00$\pm$.01\\
[0.2cm]
 {\mgi} 457.1       &  -     &   -     &  -    &   0.31 & 0.10  &  0.25  &  0.50   &  0.31 & 1.00$\pm$.32\\
 {\sii} 671.6       &  0.33  &   6.86  &  1.25 &   1.53 & 0.09  &  0.84  &  3.80   &  1.53 & 1.00$\pm$.06\\
 {\sii} 673.1       &  0.58  &  11.20  &  2.08 &   2.86 & 0.07  &  1.41  &  7.36   &  2.80 &  .98$\pm$.02\\
 {\sii} 406.9       &  0.31  &   4.92  &  0.95 &   1.60 & 0.26  &  0.51  &  4.23   &  1.38 &  .86$\pm$.14\\
 {\sii} 407.6       &  0.10  &   1.60  &  0.31 &   0.54 & 0.18  &  0.16  &  1.37   &  0.45 &  .82$\pm$.27\\
 {\siii} 631.2      &  1.51  &   2.73  &  1.68 &   1.80 & 0.06  &  1.68  &  2.25   &  1.82 & 1.01$\pm$.03\\
 {\siii} 953.1+906.9 &  58.00  & 101.00 & 64.00 &    -   &    -  & 67.60  & 84.40   & 71.50 &       -     \\
 {\cliii} 551.7      &  -     &   -    &  -    &   0.25 & 0.04  &  0.24  &  0.26   &  0.25 &  .99$\pm$.15\\
 {\cliii} 553.7      &  -     &  -     &  -    &   0.34 & 0.04  &  0.31  &  0.48   &  0.35 & 1.02$\pm$.12\\
 {\ariii} 713.6+775.1 &  -     &  -     &  -    &  11.60 & 0.40  &  8.63  & 11.80   &  9.36 &  .80$\pm$.03\\
 {\ariv} 471.1       &   -    &  -     &  -    &   2.00 & 0.12  &  3.17  &  2.06   &  2.91 & 1.46$\pm$.09\\
 {\ariv} 474.0       &   -    &   -    &  -    &   2.90 & 0.08  &  3.44  &  3.22   &  3.39 & 1.17$\pm$.03\\
 {\ariv} 717.1+726.3+ &  -     &   -    &  -    &   0.33 & 0.10  &  0.26  &  0.25   &  0.26 &  .79$\pm$.24\\
 {\kiv} 610.2+679.5   &  -     &   -    &  -    &   0.17 & 0.05  &  0.17  &  0.15   &  0.17 & 1.00$\pm$.30\\
\multicolumn{3}{l}{\underline{Line flux ratios}} & & & &&&  &\\
  {\oii} 372.9/372.6  & 0.41   &    0.42 &  0.42 &  0.44 &  0.02 &  0.45  &  0.37  & 0.42  &  0.96$\pm$.04\\
  {\sii} 671.6/673.1  & 0.56   &    0.61 &  0.60 &  0.53 &  0.03 &  0.58  &  0.50  & 0.53  &  1.02$\pm$.06\\
 {\cliii} 551.7/553.7 &   -    &   -     &  -    & 0.73  &  0.15 &  0.79  &  0.54  & 0.74  &  1.01$\pm$.20\\
 {\ariv} 471.1/474.0  &   -    &   -     &  -    & 0.69  &  0.05 &  0.92  &  0.64  & 0.86  &  1.24$\pm$.09\\
  {\nii} 658.4+/575.5  &46.40  & 48.70  & 47.70 & 42.80 &  5.00 &  51.00  & 46.80  & 47.90 & 1.12$\pm$.13\\
  {\oii} 372.6+/732.0+ & 6.50  &  6.79  &  6.65 &  6.23 &  0.20 &   8.75  &  4.88  & 6.20  & 1.00$\pm$.03\\
  {\oiii} 500.7+/436.3 & 104.00&  99.50 & 103.00& 107.00&  1.70 & 108.00  & 101.00 & 107.00& 1.00$\pm$.02\\ 
  {\sii} 671.6+/406.9+ & 2.24  &  2.78  &  2.63 &  2.05 &  0.42 &   3.35  &  1.99  & 2.37  & 1.16$\pm$.24\\
  {\ariv} 471.1+/717.1+&  -    &   -    &  -    & 14.90 &  4.60 &  25.20  & 20.90  & 24.40 & 1.64$\pm$.50\\
[0.1cm]
\hline
\end{tabular}
\end{flushleft}
\end{table*}

The NEBU model (col. 3 of Table 1 and col. 7-9 of Table 2) 
consists of two constant-pressure sectors, implying 
five free parameters: two gas pressures, one optical depth, 
one covering factor and the inner radius. 
The matter-bounded and radiation-bounded sectors 
now have angular ``diameters'' 1.31$''$ and 0.84$''$ 
respectively, in fair agreement with observation. 

In col.~10 of Table~2 are given the ratios of computed to observed fluxes 
for all lines (and the 1-$\sigma$ observational uncertainties). 
Very few of these ratios are significantly different from unity. 

The two-sector description is successful for all \nii, \oii, \sii, and \cliii\ 
lines, once a larger density, called for independently by geometrical 
information, is postulated for the short axis. The \ariv\ doublet ratio 
suggests an even denser region which may be accomodated with \nii\ and 
\sii\ ratios within uncertainties, but proved difficult to reconcile with 
the \oii\ ratio in this description. 

The \oi\ doublet is now predicted at $\sim$~60\% 
of the observation, which can be considered satisfactory. 
Departure from strict ionization/thermal equilibrium, 
quite possible in ionization fronts, may increase the \oi\ emission. 

A 50\% mismatch of the argon ionization balance 
is left as a noticeable imperfection of the model. Atomic data, 
notably the di-electronic recombination rates introduced on 
empirical grounds (Sect.~3.1.1), may be incorrect for argon and 
this discrepancy does not justify considering a more elaborate 
gas distribution. Nonetheless, if part of the 
high-ionization region is denser than the constant-pressure 
dense sector, both the \ariv\ doublet ratio 
and the \ariv\ to \ariii\ ratio can improve without much influencing 
the \oii\ ratio. This possibility was not explored. 

\subsubsection{Comments}

In the NEBU model, the ``colour temperature'' of the central star 
below the He$^+$ ionization limit is $T_B$ = 1.25$\times$10$^5$K. 
Subtracting 1.5$\times$10$^4$K from $T_B$ (Sect.~3.1.2), the 
\Teff\ of the model star derived from the NEBU calculation exceeds 
the one used in the HC calculation by 2$\times$10$^4$K. A much lower $T_B$ 
(and correspondingly larger $f_4$) results in a weaker, still acceptable 
\cii\ (reminiscent of the HC result) but the \sii\ line fluxes can 
no longer be reproduced in the NEBU computation to the quoted accuracy. 
Good results are obtained in the range 
[$T_B$ = (1.20 -- 1.35)$\times$10$^5$K, $f_4$ = (0.342 -- 0.253)]
although the upper bound to $T_B$ is 
loosely defined due to the asymptotic behaviour of intensity predictions, 
while the lower bound is not strict either considering uncertainties 
in some atomic data for sulfur. 

A major feature of model atmospheres of hot PN nuclei is the amplitude 
of the He$^+$ ionization-limit discontinuity at \la22.8. The discontinuity 
factor, as derived empirically from NEBU models of 
Wray, decreases from $f_4$ = 0.406 for \Teff\ = 1.0$\times$10$^5$K 
($T_B$ = 1.15$\times$10$^5$K) to  $f_4$ = 0.309 for 
\Teff\ = 1.1$\times$10$^5$K ($T_B$ = 1.25$\times$10$^5$K, present model), 
while the same quantity increases from 0.36 to 0.50 according 
to the standard NLTE stellar atmosphere models of 
Clegg \& Middlemass (\cite{clemid87}) 
(assuming a star mass of 0.6~\Msun, but the dependence on $g$ is moderate 
here). These determinations intersect for \Teff\ = 1.02$\times$10$^5$K, 
near to the lower end of the \Teff\ range leading to the most satisfactory 
NEBU models. Thus the Clegg \& Middlemass (\cite{clemid87}) 
results appear to provide a fair representation of 
the actual spectral energy distribution of the nucleus of Wray~16-423 
as depicted from nebula properties. 
It is frequently suspected that, in order to best account for nebular 
properties, the departure from a black-body distribution in the 
EUV continuum of hot stars should be {\sl less} than in standard NLTE 
model atmospheres (\eg, Harrington \cite{harrin89}). Also detailed 
radiation-transfer calculations with self-consistent treatment 
of wind in hydrodynamic NLTE model atmospheres show a significant rise 
in the continuum flux shortwards of \la22.8, thus partly ``erasing'' 
the discontinuity (Kudritzki \& M\'endez \cite{kudmen93}; 
Yamamoto et al. \cite{yamsel97}). Finally, shocks may develop in the 
strong winds of hot [WR] stars and produce extra emission 
(\eg, Feldmeier et al. \cite{felkud97}). In this context, 
the {\sl relatively large He$^+$ discontinuity} found to 
characterize the nucleus of Wray~16-423 might be due to 
the large helium abundance in its atmosphere. 

Considering that, for large \Teff's, (1) the ``asymptotic'' behaviour 
of the NEBU solutions becomes artificial and (2) a large \la22.8 
discontinuity is then unlikely whatever the helium content 
of the atmosphere may be, the lower end of the NEBU \Teff\ 
range must be preferred. One more aspect is the shape of the 
nebula: for large \Teff's, the model nebula is found to 
become spherical and bigger than observed. 
The final best estimate, which 
takes into account uncertainties on \sii\ atomic data as well as 
the tendency shown by the HC model, is: 

\Teff(Wray~16-423) = (1.07$\pm$0.10)$\times$10$^5$K. 

The source luminosity in the NEBU model is 92\% of that of the HC model star. 
The number of ionizing photons N$_{13.6}$ differs by $\sim$~5\% (Table~1). 
Part of this difference can be ascribed to the slight overestimate of \Ib\ 
in the HC model ($+3.6\%$ on average for \Ib\ and radio, col.~4 of Table~2). 
Moreover, the black bodies used in the NEBU computation are 
intended to represent the ionization radiation shortward of 91.2\nm, 
whereas the stellar flux is generally larger longward owing to 
the H$^0$ discontinuity: for high \Teff's the correction 
to $L_B$ from this cause is a few percent. Thus, despite the 
different nebula structures considered, both models lead to 
almost identical central star luminosities. 
The adopted luminosity is: 

\Lstar(Wray~16-423) = (4.35$\pm$0.15)$\times$10$^3$\Lsun,

\noindent
where the 3\% uncertainty from modeling is to be added 
to the 4\% observational uncertainty on \Ib. 

\subsection{He~2-436}

\subsubsection{HC computation}

The ionized mass of He~2-436 is small and it 
is likely to be radiation bounded in all directions. 
Using the oxygen abundance of Paper~I, 
the \oiii\ ratio could be matched by taking 
\Nh = 1.3$\times$$10^5$\cmc\ in the O$^{2+}$ zone, but 
the \oii\ 372.7\nm/732.5\nm\ ratio indicates a lower density. 
An outer shell was therefore added to the main shell. 
The model (col.~2 of Table~3 and results in col.~2 of Table~4) 
meets difficulties in reproducing the strengths of \oii\ and \oi. 

The model is optically thick at 1.5 and 4.9~GHz and the predicted 
fluxes agree very well with the observed fluxes at these frequencies. 
Even though the model is not yet completely thin at 8.4 GHz, 
it appears to overestimate the flux there, confirming the 
suspicion that \Fb\ has been overestimated (Sect.~3.2.2). 

The HC model helium abundance exceeds the empirical one. 
Standard empirical collisional contributions to \hei\ lines 
are larger than in this model, which allows for 
the photoionization of $2^3S$ (Clegg \& Harrington \cite{clehar89}).  
This large helium abundance may be spurious, considering the bad fit 
to the \hei\ lines. Photoionization of $2^3S$ 
may be partly inhibited by dust absorption (Sect.~4.3). 

\subsubsection{NEBU computation}

The gas distribution adopted in the HC computation cannot account for 
low-ionization lines. The importance of the low-ionization region is 
controlled by N$_{13.6}$/($r^2\times$\Nh). 
N$_{13.6}$ is determined by \Ib, \Nh\ by low-ionization line diagnostics, 
and the size $r$ of the emission region by the 20~cm continuum 
surface brightness. Increasing 
low-ionization line fluxes in the {\sl strictly spherical} model 
implies that part of the photoionized gas, not dominant 
in terms of \Hb\ emission, lies further away from the 
star than the bulk of emitting material. The radio-continuum slope 
suggests a geometrically thick nebula with density gradient (Sect.~2.2). 

The NEBU spherical model consists of two narrow shells, 
separated by a low-pressure region of substantial mass but 
negligible emission. The parameters describing the structure are 
the pressure and radius of each shell, and the 
fraction of \Ib\ emitted by the outer shell. There are thus five 
structure parameters as in the computation for Wray~16-423. 
Models could be found which matched most line 
fluxes as closely as in the case of Wray~16-423 with the 
exception of the \nii\ lines: the flux ratio 
\nii\ 658.4+4.8\nm/575.5\nm\ was systematically predicted too large 
by 40--50\%, a discrepancy much larger than the 10\% observational 
uncertainty. An example is provided by Model~M1 (Table~3, col.~3).  
Flux predictions for M1 are not given in Table~4, as they 
turn out to almost coincide with those of Model~M2 below, 
except for the \nii\ lines and the radio fluxes. 
The \oii\ and \nii\ lines cannot be made compatible in Model~M1 
because their ratios point to significantly different average densities. 
A combination of regions of similar ionization 
but greatly different densities is required. Hence low-ionization 
regions must exist at different distances from the star, 
indicating a failure of the assumption of spherical symmetry. 
Aside from this spectroscopic verdict, the structure obtained 
assuming strict sphericity is artificial in that 
the inner shell absorbs most of the photons: a moderately 
larger column density would suffice to exhaust the ionizing 
radiation. An inner shell, thick along 
some directions and thin along other ones, is more realistic. 
Another difficulty with spherical models is that the shells 
must be totally disconnected from each other: absorption by 
low-density gas in between the shells would produce 
high-ionization lines and reduce the amount of radiation 
available in low-ionization regions. 

\begin{table}
\caption[]{Models for He~2-436}
\begin{flushleft}
\begin{tabular}{lccc}
Model                         &   HC   &   M1    &  M2     \\
Distance/kpc                  &   25   &   25    &  25      \\ 
(\Teff\ or $T_B$)/10$^3$K     &   70   &   80    &  85      \\
(\Lstar\ or $L_B$)/10$^3$\Lsun  &  6.90  & 5.06$^a$ & 5.12$^b$ \\  
log~$g$ \ \ \ \ \ \ \ \ ($f_4$) &   4.7  & (.10) &  (.08)   \\
N$_{13.6}$/10$^{47}$s$^{-1}$  & 4.82   &  3.76   &   3.76   \\
Filling factor               & 1.0 &  1.0 & 1.0   \\
Mass/10$^{-2}$\Msun\         & 1.9 &  2.0 & 3.8  \\
\multicolumn{3}{l}{Radius/10$^{-3}$pc:}  &   \\
(Inner shell)                &  5.3-9.5   & 7.3-8.4   & 8.3-9.0  \\
(Outer shell)                &  9.5-11.0  & 18.4-19.3 & 8.3-31.  \\
\multicolumn{3}{l}{\Nh/10$^4$\cmc\ \ (and covering factor):}  \\
(Inner)                      & 13. &  22. & 27. (.62)   \\
(Outer)		             & 6.0 &  6.4 & 2.0 (.38)  \\
\multicolumn{3}{l}{Gas pressure/10$^{-8}$CGS:}  \\
(Inner)                     &  -   &  75. &  85.    \\
(Outer)                     &  -   &  23.  &  7.    \\
\multicolumn{3}{l}{Abundances by number:$^c$}   & \\
H                        &  1.00 &  1.00 &   1.00 \\
He                       &  0.12 & 0.104$^d$ & 0.104$^d$ \\
C \  ($\times$10$^{5}$)  &  150. &  110. &  111.  \\
N \  ($\times$10$^{5}$)  &  3.50 &  2.34 &   2.64 \\
O \  ($\times$10$^{5}$)  &  21.0 &  24.5 &  23.1 \\
Ne ($\times$10$^{5}$)    &  3.16 &  3.54 &   3.42 \\
S \ \ ($\times$10$^{5}$) &  0.50 & 0.419 &   0.391 \\
Ar ($\times$10$^{5}$)    &  --   & 0.060 &   0.060 \\
\end{tabular}

\ \ $^a$ 4.82, correcting for $f_4$ (see text). \\
\ \ $^b$ 4.85, correcting for $f_4$ (see text). \\
\ \ $^c$ Mg, Cl and K as in Wray~16-423 (Table~1). \\
\ \ $^d$ Adopted He abundance will be 0.108 (Sect.~4.3). \\
\end{flushleft}
\end{table}

\begin{table*}
\caption[]{Observations and model predictions for He~2-436}
\begin{flushleft}
\begin{tabular}{lr|rr|rrrr}
           & \multicolumn{1}{c}{HC} \vline  & 
\multicolumn{2}{c}{Observation} \vline & \multicolumn{4}{c}{NEBU}  \\
\cline{2-8}
	& \multicolumn{1}{c}{Model} \vline & Flux & 1$\sigma$ Err & High \Nh & Low \Nh & Model M2 & M2/O \\
Covering factors:              &       &       &        &  0.624 &  0.376 & 1.000  &  -         \\ 
\multicolumn{1}{l}{\underline{Absolute fluxes}}  &&&&&&& \\
log\Ib(erg cm$^{-2}$s$^{-1}$)  & -11.57 & -11.59 & 0.06 & -11.65 & -11.65 & -11.65 & 0.88$\pm$.13 \\
1.46 GHz (mJy)                 & 0.5   & 0.62   & 0.25  &  0.32  &  3.10  & 1.36$^a$ & 2.20$\pm$.90 \\
4.89 GHz (mJy)                 & 3.7   & 3.90   & 0.20  &  2.97  &  6.81  & 4.41$^a$ & 1.13$\pm$.06 \\
8.40 GHz (mJy)                 & 6.1   & 4.90   & 0.20  &  4.95  &  6.89  & 5.68$^a$ & 1.16$\pm$.05 \\
\multicolumn{3}{l}{\underline{Relative line fluxes} (wavelengths in \nm)}  &&&&& \\
 \hi\ 486.1              & 100.00 & 100.00 &    -   & 100.00   & 100.00   & 100.00   &        -      \\
 Cont. 364.2 (/\nm)      &    -   &  43.6  & 3.50   &   44.7   &   48.5   &   46.1   & 1.06$\pm$.08  \\
 Cont. 364.8 (/\nm)      &    -   &  6.80  & 0.70   &   6.00   &   8.90   &   7.10   & 1.05$\pm$.10  \\
 \hi\ 656.3              & 288.00 & 288.00 & 1.6    & 289.00   & 289.00   & 289.00   & 1.00$\pm$.01  \\
 \hei\ 447.1             & 5.85   & 6.07 &  0.21    & 6.07     & 5.93     & 6.02     &  .99$\pm$.03  \\ 
 \hei\ 587.6  		& 16.00  &18.50 &  0.17    & 18.90    & 18.20    & 18.60    & 1.01$\pm$.01  \\ 
 \heii\ 468.6            & 0.16    & $<$0.7 &   -   & 0.47     & 0.48     & 0.48     &   $>$ 0.64    \\
 \cii\ 426.7  		& 0.79    & 0.96   & 0.22  & 0.97     & 0.87     & 0.93     & .97$\pm$.22  \\ 
 \ciiip\ 418.7           &   -    & $<$0.50 & 3$\sigma$ &  0.07 & 0.14    & 0.10     & $>$ 0.20  \\
 \ciii\ 190.9+0.7		& 1210.00 &   -    &  -    &1150.00   & 912.00   &1060.00   &        -   \\ 
 \civ\  154.9 		&  715.00 &   -    &  -    & 166.00   & 233.00   & 191.00   &        -   \\ 
[0.2cm]
 {\nii} 658.4+4.8		& 7.67   &14.30  & 0.75     &  9.01    & 22.20    & 14.00    &  .98$\pm$.05  \\ 
 {\nii} 575.5		& 0.54   &1.11   & 0.09     & 1.49     & 0.57     & 1.15     & 1.03$\pm$.08   \\
 {\oi} 630.0+6.3 	        & 0.88   & 6.83  & 0.09     & 4.08     & 3.73     & 3.95     &  .58$\pm$.01  \\
 {\oii} 372.6   		& 3.29   & 8.25  & 0.90      & 2.82    & 19.90    & 9.23     & 1.12$\pm$.12  \\
 {\oii} 372.9   		& 1.01   & 3.08 & 0.86      & 0.84     & 6.66     & 3.02     &  .98$\pm$.27   \\
 {\oii} 732.0+3.0		& 5.93   & 13.60 & 0.12     & 15.90    & 9.61    & 13.50    &  .99$\pm$.01  \\
 \oiip\ 465.1+           &   -    & $<$5.00 & -     &    0.26   &  0.27   & 0.27     & $>$ 0.05\\
 {\oiii} 500.7+495.9 	&1090.00 &1065.00& 6.00     & 959.00  & 1237.00  & 1063.00  & 1.00$\pm$.01  \\
 {\oiii} 436.3   	& 14.00  & 14.20 & 0.30     & 16.80    &  9.76    & 14.20    & 1.00$\pm$.02  \\ 
 {\neiii} 386.8+396.7	& 77.00  & 75.30 & 1.70     & 77.90    & 71.00    & 75.30    & 1.00$\pm$.02  \\
[0.2cm]
 {\mgi} 457.1            &  -     & $<$0.39 & 3$\sigma$ & 0.35  &   0.23  &   0.30  & $>$ 0.79  \\
 {\sii} 671.6	        & 0.39   & 0.50   & 0.03     & 0.14    & 1.14    & 0.51    & 1.03$\pm$.06  \\
 {\sii} 673.1		& 0.82   & 1.13   & 0.03     & 0.32    & 2.39    & 1.10    &  .97$\pm$.03  \\
 {\sii} 406.9            & 2.24   & 2.36   & 0.55     & 3.28    & 2.33    & 2.93    & 1.24$\pm$.29  \\
 {\sii} 407.6            & 0.69   & 0.82   & 0.43     & 0.99    & 0.76    & 0.90    & 1.10$\pm$.57  \\
 {\siii} 631.2		& 1.10   & 2.52   & 0.04     & 3.16    & 1.57    & 2.57    & 1.02$\pm$.02  \\
 {\siii} 953.1+906.9      & 68.00  &   -    &    -     & 62.50   & 66.20   & 63.90   &          -    \\
 {\cliii} 551.7          &  -    & $<$0.24 & 3$\sigma$ & 0.03  &  0.13   & 0.07    & $>$ 0.27  \\
 {\cliii} 553.7          &  -    & $<$0.24 & 3$\sigma$ & 0.12  &  0.34   & 0.20    & $>$ 0.84  \\
 {\ariii} 713.6+775.1     &  -    &   8.12  & 0.30    &  8.70   &  6.44   &  7.85   &  .97$\pm$.04  \\
 {\ariv} 471.1           &  -    &   0.22  &  0.19   &  0.15   &  0.79   &  0.39   & 1.79$\pm$1.5  \\
 {\ariv} 474.0           &  -    &   0.74  & 0.14    &  1.02   &  1.87   &  1.34   & 1.81$\pm$.35  \\
 {\ariv} 717.1+726.3+     &  -    & $<$0.35 & 3$\sigma$&  0.20  &  0.13  &  0.17  & $>$ 0.48  \\  
 {\kiv} 610.2+679.5       &  -    & $<$0.24 & 3$\sigma$&  0.07  &  0.05  &  0.06  & $>$ 0.25  \\  
\underline{Line flux ratios} & & & & & &  \\
   {\oii} 372.9/372.6      & 0.31  & 0.37  & 0.11   &0.30  &0.33  &0.33  & 0.87$\pm$.27  \\
   {\sii} 671.6/673.1      & 0.48  & 0.44  & 0.03   &0.43  &0.47  &0.47  & 1.06$\pm$.07  \\
  {\ariv} 471.1/474.0      &  -    & 0.29  & 0.24   &0.14  &0.42  &0.29  & 1.00$\pm$.80  \\
  {\nii}  658.4+/575.5     & 14.20 & 12.90 & 1.20   &6.04  &38.90 &12.20 &  .94$\pm$.09  \\
   {\oii}  372.6+/732.0+   & 0.73  & 0.83  & 0.13   &0.23  &2.77  &0.91  & 1.09$\pm$.16  \\
  {\oiii} 500.7+/436.3     & 77.60 & 75.00 & 1.60   &57.00 &127.00&75.00 & 1.00$\pm$.02  \\
   {\sii}  671.6+/406.9+   & 0.20  & 0.51  & 0.16   &0.11  &1.14  &0.42  &  .82$\pm$.26  \\
[0.1cm]
\hline
\end{tabular}

\ \ $^a$ No shielding between sectors included (see Sect.~3.4.2). \\
\end{flushleft}
\end{table*}

Two-sector radiation-bounded models were built, with only four parameters 
describing the gas distribution: two pressures, one covering factor 
and the inner radius. In accordance with Sect.~3.2.2, 
the absolute \Hb\ flux is taken as only 87\% of the flux quoted 
by Webster (1983). Despite reduced freedom, the two-sector model 
(Model M2: col.~4 of Table~3 and cols.~5-7 of Table~4) is more 
successful than any spherically symmetric model. 

Nonetheless the low-frequency radio flux (1.5~GHz) is 
discrepant. The total radial optical depth at 1.5~GHz 
is large ($\sim$~8), but the outer ``radius'' 
of the low-density sector (radial optical depth $\sim$~1)
is also large in order to account for the strength of 
those low-ionization lines easily quenched by collisions. 
Now, radio fluxes given in col.~7 of Table~4 are rather 
{\sl upper limits} as they result from simply averaging the 
individual fluxes, that is without considering possible ``shielding'' 
effects between sectors. 
Given that self-absorption is important and spherical symmetry is 
abandoned, geometry must be considered. The nebula may, 
for example, be elongated in the direction of the observer, 
then reducing somewhat the 1.5~GHz flux. A larger reduction 
would obtain if a smaller solid angle were given to the 
low-density sector but some lines would then not be nearly 
as well matched as in Model~M2. At this stage it is worthwile to note that 
(1) this two-sector model is a coarse representation and 
(2) some atomic data for ionization equilibria used in judging 
the model fit (\eg, sulfur) are not of ultimate accuracy. 
Density may be larger in part of the inner layers of 
the otherwise ``low-density'' sector, then reducing the outer radius and 
the 1.5~GHz flux. 

The high-frequency radio fluxes (4.9 and 8.4~GHz) are 10--20\% too large 
in Model~M2, but the radio continuum slope is matched, 
suggesting that \Ib\ may be even less than the lower limit of  
the error bar. However, aspect effects in the incomplete  
inner shell introduce some uncertainty in 
the computed flux: the 8.4~GHz optical depth across this 
component is only 0.19 radially, but 0.85 near to 
the edge. The uncertainty of geometrical origin amounts to $\sim$~10\% 
and Model~M2 can be considered as compatible with both the 
optical and high-frequency radio fluxes within errors. 

As for the fit to the optical spectrum (relative to \Hb), Model M2 for 
He~2-436 is as satisfying as model M for Wray~16-423 
and the two discrepancies left in M2 are very reminiscent of 
those already noted in the case of Wray~16-423, namely M2/O = 0.58$\pm$0.01 
compared to M/O = 0.56$\pm$0.02 for \oi\ 630.0+6.3\nm\ 
and 1.9$\pm$0.4 compared to 1.6$\pm$0.1 
for the ratio \ariv471.1+4.0/\ariii713.6+775.1.  

\subsubsection{Comments}

Due to observation uncertainties (\oii\ 372.7, \sii\ 406.9, 
\cii\ 426.7, \ariv\ 471.1) and partial collisional quenching of \oiii, 
a range of two-sector NEBU models, in which all parameters 
were coherently fine-tuned, could be obtained. Model M2 is typical. 
After correcting for $f_4$, the fluctuation of black-body luminosity 
$L_B$ among the NEBU models is $\sim$ 1\% and $L_B$ is $\sim$ 1.42 times 
less than \Lstar\ (HC model). The ionizing-photon output N$_{13.6}$ 
is 1.28 times less than in the HC model. The differences between the HC 
and NEBU results can be understood as follows. 

Note first that the most significant radio flux in terms of 
photon counting, namely the 8.4~GHz flux, is predicted by the HC model 
as 1.25$\pm$0.05 times the observed value. Reaching agreement would require 
lowering N$_{13.6}$ by a factor over 1.3, as the radio optical 
depth decreases together with the column density of ionized material. 
The HC model corresponds to the original dereddened flux \Ib(c=0.61), 
the NEBU model to 0.87$\times$\Ib(c=0.58), hence a factor 1.20 smaller. 
The remaining factor 1.28/1.20 = 1.06 on N$_{13.6}$ corresponds 
to the larger fraction of photons removed from the ionizing field by 
He$^0$ photoionisation in the HC model, due mainly to the larger 
helium abundance. 

The same factor 1.28 applies to \Lstar(HC), but a supplementary 
factor arises from the 
fact that model stellar atmospheres for \Teff$\sim$7$\times$10$^4$K show 
quite a large discontinuity at 1~Ryd, implying an upward correction 
to $L_B$. After removing the 
effects of considering different photoionization rates for He(2$^3$S) 
and black bodies instead of star atmospheres, the differences 
between the HC and NEBU central star luminosities essentially reflect 
a different emphasis on which absolute flux should be favoured. 
Favouring the more accurate 8.4 GHz flux, while respecting the 
error bars on \Ib, is taken as the best option. In 
conclusion, the best estimate for \Lstar\ is the value 
obtained in the HC computation, divided by 1.28, that is: 

\Lstar(He~2-436) = (5.40$\pm$0.40)$\times$10$^3$\Lsun,

\noindent
where the 7\% uncertainty is from models only, to be added to the 7\% 
observational uncertainty on the high-frequency radio continuum fluxes. 
The model uncertainty on \Lstar\ is substantial, considering that the 
connection between \Lstar\ and the ionizing flux is not 
straightforward and that, due to the dependence on geometry, 
even a fully accurate 8.4 GHz flux does not imply a unique 
value for \Lstar\ (Sect.~3.4.2). 

If the luminosity of the central star of He~2-436 is not very 
accurately determined from models and high-frequency radio data, 
the effective temperature seems to be. 
The most satisfactory NEBU models indicate 
$T_B$ = (0.85$\pm$0.05)$\times$10$^5$K, 
which converts (Sect.~3.1.2) into: 

\Teff\ = (0.70$\pm$0.05)$\times$10$^5$K, 

\noindent
in agreement with the HC result (Table~3, col.~2). 

Some freedom on star parameters is currently unavoidable, given 
the presence of quite prominent Wolf-Rayet features in the spectrum of 
He~2-436 and the questionable accuracy of stellar atmosphere models 
for this class of stars (\eg, Crowther et al. \cite{cropas99}). 
If the \la22.8 discontinuity is large, perhaps as a 
consequence of the large helium abundance in the atmosphere, relatively 
high \Teff's are possible. If, on the other hand, this discontinuity 
is ``erased'' by the emission of a strong hot wind, then relatively low 
\Teff's may have to be preferred. Considering these additional 
uncertainties, our estimate for the He~2-436 star effective 
temperature is, more conservatively: 

\Teff(He~2-436) = (0.70$\pm$0.10)$\times$10$^5$K.  

A more definite picture should await more accurate fluxes for 
some optical lines (see above) and new spectroscopic diagnostics. 
The He~2-436 models may be distinguished thanks to 
their UV lines (Table~4). In models with relatively cool 
central stars, the carbon lines are weak. 
In other models, these lines are as strong as \oiii. 

\section{Discussion}

\subsection{Appraisal of the models} 

The two codes used to model the two Sagittarius PNe reached agreement under 
similar assumptions, confirming the reliability of the calculations. 
A few differences, notably on sulfur, 
were understood in terms of atomic data updating. The HC approach 
was restricted to simple assumptions and provided the basis for the more 
systematic approach adopted with NEBU. The HC and NEBU models led to similar 
properties for both the star and the chemical composition, but 
some significant differences demonstrated the influence of the adopted 
gas distribution. In the case of Wray~16-423, 
the 33 independent observables available (1 imagery; 3 photometry; 
29 relative optical spectroscopy, out of which 18 and 11 
are accurate to better than 4\% and 2\% respectively) against the 
18 free parameters of the NEBU model (1 reddening; 10 abundances; 
2 stellar; 5 nebular parameters) lead to some 15 ``redundancies'' 
to estimate the model consistency. 

Correspondingly in He~2-436, 27 observables 
(4 photometry; 23 relative optical spectroscopy, out of which 
12 and 8 are accurate to better than 4\% and 2\% respectively) against 
14 parameters (1~reddening; 7 abundances; 2 stellar; 4 nebular 
parameters), give 13 redundancies. 

Except for \ariv\ and \oi, all spectroscopic features are explained to 
1~$\sigma$ by the NEBU models of both PNe. 
The discrepancies for either \ariv\ or \oi\ are strikingly similar 
in both objects (end Sect.~3.4.2), suggesting they have 
systematic origins. The mismatch to \oi\ has no effect on the 
oxygen abundance determination.

The overestimation  of the radio flux of He~2-436 
may result from not taking into account the mutual 
shielding of the two sectors and possible aspect effects 
(Sect.~3.4.2). 

In conclusion, physically independent lines from different elements and 
ions, involving redundant information, could be made to fit 
into self-consistent pictures. 
Although the model for He~2-436 is not really unique (Sect.~3.4.2) 
and the model for Wray~16-423 may be too coarse a description 
of the innermost regions (Sect.~3.3.2), 
these models appear as valuable descriptions of the Sagittarius PNe, 
giving a high degree of confidence in the accuracy of the derived 
elemental abundances and central star properties. 

\subsection{Geometry of Wray~16-423 and He~2-436} 

Both PNe were described in terms of two-sector models, but the 
geometrical interpretation must be different. 
The moderate contrasts in density and radius of the two sectors 
of Wray~16-423 suggest an ellipsoidal structure, also indicated 
by radio and optical imaging. The short axis is denser and 
radiation bounded as in many relatively young Galactic PNe. 
The large contrasts in density ($\sim$15) and thickness ($\sim$30)
found for the sectors of the He~2-436 model suggest a two-shell 
geometry in which the outer shell is illuminated by radiation leaking 
through ``holes'' in a very dense radiation-bounded incomplete 
inner shell. 

The ionized mass of the inner shell of He~2-436 is only 
$\sim$ 4$\times$10$^{-3}$~\Msun. This shell could correspond to a 
recent episode of mass ejection resulting from a 
late thermal pulse during the early post-AGB phase 
(\eg, Sch\"onberner \cite{schoen97}).  
With time, this shell may 
rapidly grow optically thin and be barely detectable if its total 
mass is small, then resembling the small amount of dense optically 
thin gas suggested by the \ariv\ ratio in Wray~16-423. 

The mass of He~2-436, obtained 
from the ionized gas and completing the outer shell assuming 
spherical symmetry, is about a third of the mass of Wray, consistent 
with He~2-436 being still radiation bounded in all directions. 
It will be argued in Sect.~4.7 that these PNe are 
quite similar objects, despite their very different appearance.

\begin{table*}
\caption{Helium lines$^a$}
\begin{flushleft}
\begin{tabular}{l|rrr|rrrr|l}
\hline 
 & \multicolumn{3}{c}{Wray~16-423} \vline & \multicolumn{4}{c}{He~2-436} \vline & \\
 & \multicolumn{3}{c}{He/H $= 0.1074$} \vline & \multicolumn{4}{c}{He/H $= 0.1041$ (M) and 0.1072 (M')} \vline & \\
 \la\ (\nm)& O  &    M &   M/O      &    O &  M & M/O  & M'/O & Comments$^b$\\
\hline
\multicolumn{6}{l}{\hei\ triplet (with 2$^3$S $- n^3$P self-absorption)} &&& \\
706.52&    7.31 &  9.77 & 1.340$\pm$.015 &  13.18 & 12.81 & .972$\pm$.016 & .912$\pm$.015 &  \\
388.86&    7.97 &  8.00 &  .996$\pm$.034 &   7.59 &  7.59 &1.000$\pm$.070 &1.016$\pm$.070 & \hi\ subtr.\\
587.57&   16.37 & 16.33 &  .998$\pm$.006 &  18.48 & 18.63 &1.009$\pm$.009 &1.000$\pm$.009 & \\
471.32&     .55 &  1.05 &(1.910$\pm$.160) &  1.26 &  1.25 &1.002$\pm$.130 &1.052$\pm$.140 & \ariv\ subtr. using $\lambda$474.0\nm\\
447.15&   5.44  &  5.45 & 1.003$\pm$.043 &   6.07 &  6.02 & .993$\pm$.034 &1.001$\pm$.034 & \ \ and predicted doublet ratio \\
412.08&$<$ .65  &   .36 & $>$ .560       &    .29 &   .42 &1.420$\pm$.300 &1.370$\pm$.300 & \\
402.62&   2.72  &  2.46 &  .904$\pm$.084 &   3.41 &  2.68 & .790$\pm$.135 & .798$\pm$.135 &\heii\ and weak \hei\ subtr.\\
381.96&   1.27  &  1.34 & 1.055$\pm$.210 &   1.19 &  1.46 &1.230$\pm$.400 &1.240$\pm$.400 & \\
370.50&   0.61  &  0.82 & 1.340$\pm$.600 &     -- &   --  &  --           & --            & \hi\ subtr.\\
\multicolumn{4}{l}{\hei\ singlet (Case B)} \vline &&&&& \\
728.13&    .82  &  1.14 & 1.392$\pm$.073 &   1.10 &  1.34 &1.216$\pm$.028 &1.172$\pm$.028 &Case A $\sim$ B/1.8\\
667.82&   3.99  &  3.97 &  .995$\pm$.043 &   4.57 &  4.41 & .964$\pm$.011 & .977$\pm$.011 & \\
504.77&    .21  &   .23 & 1.093$\pm$.270 &$<$ .25 &   .26 &     $>$ 1.040 &     $>$ 1.020 &Case A $\sim$ B/1.4\\
396.47&    .54  &  1.32 & 2.440$\pm$2.00 &    .86 &  1.41 &1.630$\pm$2.00 &1.670$\pm$2.00 &Case A $\sim$ B/30, \neiii\ subtr.\\
492.19&   1.33  &  1.34 & 1.010$\pm$.052 &   1.64 &  1.47 & .896$\pm$.070 & .915$\pm$.071 & \\
443.76&$<$  .21 &   .08 & $>$ .360       &$<$ .23 &   .08 &     $>$  .370 &     $>$  .360 &Case A $\sim$ B/1.3\\
361.36&$<$ 1.37 &   .64 & $>$ .470       &     -- &   --  &  --           & --           &Case A $\sim$ B/25 \\
438.79&    .74  &   .62 &  .839$\pm$.113 &    .50 &   .67 &1.340$\pm$.300 &1.370$\pm$.300 & \\
\hline
\end{tabular} 

$^a$ Observed dereddened fluxes (O) and model fluxes (M) in units \Hb\ = 100.\\
$^b$ ``Case~A $\sim$ B/2'' means that, under Case~A conditions, the predicted line intensity would be half the Case~B value.

\end{flushleft}
\end{table*}

\subsection{Helium abundance}

A number of \hei\ lines were observed in the Sagittarius PNe 
(Paper~I), providing a way to check the consistency of the 
data sets used (theory and observation) and the reliability 
of the helium abundance determination. 

In the NEBU models, all lines from levels $n < 6$ of He$^0$ 
are computed using recent effective recombination coefficients 
(Smits \cite{smits96}; Smits, 1999, private communication) and 
collision strengths (Sawey \& Berrington \cite{sawber93}) and accurate 
radiative transition probabilities (Kono \& Hattori \cite{konhat84}). 
New fitting formulae are accurate to typically 0.2\% 
(maximum error $\sim$~1\%, generally at a few 10$^2$~K). 
Case~B recombination is assumed for all singlet lines but 
2$^3$S $- n^3$P self-absorption is taken into account. 

2$^3$S $- n^3$P self-absorption and fluorescence are treated in a 
semi-empirical local-escape formalism: the escape formula is 
taken as a guideline to connect the different probabilities 
(for the different transitions at the different locations in 
the nebula) and a constant velocity width is {\sl chosen} to account 
for $\lambda$388.8 (2$^3$S $- 3^3$P). 

In col.~1 of Table~5 are given the wavelengths for 
9 triplet and 8 singlet lines of \hei. The observed 
de-reddened intensities and upper limits (``O'', Paper~I 
with modified reddening correction for He~2-436, Sect.~3.2.2), 
the model intensities (``M'', NEBU models, Table~2 and Table~4) 
and their ratios (``M/O'', with one more exact digit) are given in 
cols. 2-4 and cols. 5-7 of Table~5 for Wray~16-423 and He~2-436 
respectively (concerning col.~8, ``M'/O'', see below). 
Subtracted blends are noted in col.~9 (``Comments'') of 
Table~5. 1-$\sigma$ errors are attached to the M/O values. 

In Table~5, the three strongest \hei\ lines, predominantly excited by 
recombination and not much sensitive to departure from Case~B 
or self-absorption ($\lambda$$\lambda$ 587.6, 447.1, 667.8), 
agree remarkably well with each other in both PNe. 
($\lambda$667.8, accurately measured in 
He~2-436, is further considered below). 
Next $\lambda$402.6\ and $\lambda$492.2 are fairly well 
explained, although $\lambda$402.6 would call for some extra 
excitation: this line is the strongest one from an $n = 5$ level 
and collisional excitation rates may not be accurate for $n > 4$ 
(Sawey \& Berrington \cite{sawber93}). 

$\lambda$471.3\ was de-blended from \ariv~471.1\nm\ using the \ariv\ doublet 
ratio from models (instead, in Tables~2~\&~4, 
the theoretical \hei~471.3\nm\ from col.~3 \& 6 of Table~5 was 
subtracted from $\lambda$471.2). The large M/O for 
$\lambda$471.3\nm\ in the case of Wray~16-423  (Table~5) is likely to 
arise from a problem with \ariv\ (as assumed in Sect.~3.3.2). Indeed 
\hei~471.3\nm\ is correctly predicted in the case of He~2-436, 
where \ariv~471.1\nm\ contributes little to the blend: this is 
interesting as $\lambda$471.3 is most enhanced by collisions. 
$\lambda$706.5, mainly excited by collisions from 2$^3$S, is 
correctly predicted in He~2-436, but not in Wray~16-423, 
where the 2$^3$S population may therefore be incorrect. 

Photoionization of 2$^3$S by the diffuse field 
is dominated by resonance lines (chiefly \La), 
whose radiation density is sensitive to 
dust absorption (Clegg \& Harrington \cite{clehar89}). 
The IRAS flux from He~2-436 is large, suggesting that the \La\  
radiation is substantially absorbed in this nebula. 
In standard NEBU models, the 2$^3$S photoionization by \La\ 
was ignored. This should be a good approximation for Wray~16-423 
but possibly not for He~2-436. 

A new He~2-436 model similar to M2 
(Sect.~3.4.2 and cols. 5-8 of Table~4), noted M2', was run assuming 
a uniform photoionization rate for 2$^3$S by the diffuse field. 
Clegg \& Harrington (\cite{clehar89}) note that the 
photoionization rate of 2$^3$S may not exceed 10-15\% of the 
collisional destruction rate. In M2', the uniform photoionization 
rate amounts on average to $\sim$~6\% and 50\% of the destruction 
rate for the dense and dilute sectors respectively: in 
this way part of the diffuse field produced in the incomplete 
inner shell is implicitly not processed locally 
but escapes into the outer shell where it adds to 
the weaker local diffuse field. The actual photoionization rates 
are not expected to be much larger than these values. 
In the new solution, the carbon abundance is 5\% larger and the 
helium abundance increases from 0.1041 to 0.1072. 
The new ratios M'/O of predicted to 
observed \hei\ line fluxes are given in col.~8 of Table~5. If 
$\lambda$706.5\ is now somewhat underestimated, the overall 
agreement turns out to be even better than in the standard case. 
The relative difference between $\lambda$667.8 and $\lambda$587.6 
is decreased from 0.045$\pm$0.020 to 0.023$\pm$0.020. Thus the 
large photoionization rate for 2$^3$S, expected on general grounds 
in the conditions of He~2-436, is to some extent confirmed 
by the improved fit to observations. 

Case~A intensities (no 1$^1$S$ - n^1$P photon degraded into non-resonant 
\hei\ photons; Brockelhurst \cite{brocke72}) are given in comments 
of Table~5. 
Departure from Case~B is likely for \hei\ singlet lines as 
resonance photons are destroyed by H$^0$ photoionisation in the 
H$^+$ region or escape at the PN boundary. Indeed M/O is greater than unity 
for $\lambda$728.1 in both PNe (Table~5). The departure from Case~B 
appears larger in Wray~16-423, in qualitative agreement with the fact that 
this PN is less radiation bounded. 
Other lines sensitive to the Case~B assumption are very weak and/or 
badly blended. $\lambda$501.5, lost in the \oiii\ wing, 
is not listed. \hei~396.5 is extracted from \neiii396.8 using the 
\neiii\ doublet ratio 
(\neiii\ is 95-98\% of the blend) with the right order of magnitude: 
the $\lambda$396.5\nm\ flux (0.54$\pm$0.44 and 0.86$\pm$1.00 for 
Wray~16-423 and He~2-436 respectively) loosely confirms a 
departure from Case~B, with again a possibly larger effect in 
Wray~16-423 than in He~2-436. 

The best helium abundance turns out to be 
0.1076 by number in both PNe, with estimated uncertainties 1.5\% 
and 2.5\% for Wray~16-423 and He~2-436 respectively. 

\subsection{Abundances}

\begin{table*}
\caption{Comparison of abundances$^\ast$}
\begin{flushleft}
\begin{tabular}{lrrrrrcccc}
\hline 
 & \multicolumn{2}{c}{Wray~16-423} &\multicolumn{2}{c}{He~2-436} & \multicolumn{1}{c}{Galactic} & 
\multicolumn{1}{c}{Solar$^b$$ \odot$} & Wray -- $\odot$ & He~2 -- $\odot$ & GPNe -- $\odot$ \\
Elem. & W97$^c$ & \multicolumn{1}{c}{Models} & W97$^c$ & \multicolumn{1}{c}{Models} & 
\multicolumn{1}{c}{PNe$^a$} & \multicolumn{1}{c}{AG89/GS98} &  &  &  \\
\hline
H  & 12.00 & 12.00$\pm$.00 & 12.00 & 12.00$\pm$.00 &12.00$\pm$.00 &
12.00 &- &- &- \\ 
He & 11.03 & 11.03$\pm$.01 & 11.02 & 11.03$\pm$.01 &11.05$\pm$.03 &
10.99 & \  0.04 & \ 0.04 & \ 0.06 \\ 
C  &  8.83 &  8.86$\pm$.06 &  9.18 &  9.06$\pm$.09 & 8.81$\pm$.30 & 
8.60/8.52  &    0.26/0.34 &   0.46/0.54 & \ 0.21/0.29 \\
N  &  7.62 &  7.68$\pm$.05 &  6.97 &  7.42$\pm$.06 & 8.14$\pm$.20 & 
8.05/7.92  & --.37/--.24 &  --.63/--.50 & \ 0.09/0.22 \\
O  &  8.31 &  8.33$\pm$.02 &  8.29 &  8.36$\pm$.06 & 8.69$\pm$.15 & 
8.93/8.83  & --.60/--.50 &  --.57/--.47 & --.24/--.14 \\
Ne &  7.50 &  7.55$\pm$.03 &  7.57 &  7.54$\pm$.06 & 8.10$\pm$.15 & 
8.09/8.08  & --.54/--.53 &  --.55/--.54 & \ 0.01/0.02 \\
Mg &  -    &  6.98$\pm$.30 &   -   &     -         &    -         & 
7.58       & --.60: &   -  &  -    \\
S  &  6.48 &  6.67$\pm$.04 &  6.30 &  6.59$\pm$.05 & 6.91$\pm$.30 & 
7.24$\pm$.06  & --.57 &  --.65 & --.33 \\
Cl &  -    &  4.89$\pm$.18 &   -   &     -         &    -         & 
5.28       & --.39:  &   - &    -   \\
Ar &  5.88 &  5.95$\pm$.07 &  5.76 &  5.78$\pm$.08 & 6.38$\pm$.30 & 
6.56/6.40  & --.61/--.45 &  --.78/--.62  & --.18/--.02 \\
K  &  -    &  4.65$\pm$.22 &   -   &     -         &    -         & 
5.13       & --.48: &   -  &   -  \\
\hline
\end{tabular} 

$^\ast$ Abundances are given on a logarithmic scale where H = 12\\
$^a$ Mean composition and scatter for non-Type I Galactic PNe 
(Kingsburgh \& Barlow \cite{kinbar94})\\
$^b$ AG89 = Anders \& Grevesse (\cite{andgre89}), 
except for C (Grevesse et al. \cite{grelam91}); 
GS98 = Grevesse \& Sauval (\cite{gresau98})\\
$^c$ Walsh et al. (\cite{waldud97}), Paper~I\\
\end{flushleft}
\end{table*}

In Table~6 are listed the empirical (cols.~2 \& 4; Paper~I) 
and model (cols.~3 \& 5) abundances determined 
for the Sagittarius PNe. The model abundances are based on the 
abundances listed in Table 1 and Table 3, giving a larger weight 
to the NEBU abundances (Sect.~4.1). The errors 
on the abundances reflect the different model determinations 
and the range of acceptable solutions found in the course of 
the NEBU exploration (see also P\'equignot et al. \cite{peqzij00}). 
Larger errors were adopted for those abundances based on weak 
lines from only one ionic stage (Mg, Cl, K). The empirical 
He, C, O, Ne and Ar abundances of Paper~I are generally confirmed. 
The sulfur abundance is increased in both PNe, especially in He~2-436. 
The nitrogen abundance is strongly increased in He~2-436, not in Wray~16-423. 

The abundances are compared to a mean for non-Type~I Galactic PNe 
(col.~6 of Table~6; Kingsburgh \& Barlow \cite{kinbar94}) and solar 
abundances. Since some of the (``preliminary'') standard 
solar abundances listed by Grevesse \& Sauval (\cite{gresau98}) 
are marginally discordant from those generally taken for years 
as ``standard'' (Anders \& Grevesse \cite{andgre89}) and 
the differences may be sufficient to influence the interpretation 
of the PN results, both sets of solar abundances are listed 
(col.~7 of Table~6) and the uncertainties attached 
to them are omitted, except in the case of sulfur for which the 
photospheric and meteoritic determinations 
were exchanged, keeping by chance the weighted 
mean unchanged. Columns $8-10$ of Table~6 provide the logarithmic 
differences between the model Sagittarius PN or empirical 
Galactic PN abundances on the one hand and the different solar 
abundances on the other hand. 

The enrichment in helium of 0.04dex relative 
to solar is much larger than expected from first dredge-up at the base 
of the giant branch (\eg\ Boothroyd \& Sackmann \cite{boosac99}), 
considering that the initial abundance could be $\sim$ 0.085 
to correspond to the low-$Z$ environment. 
There is no clear explanation for this large helium abundance. 
If this were a consequence of third dredge-up, a correlation with 
carbon abundance would be expected, but this is not observed.  

The mass of carbon brought to the surface 
of the star by third dredge-up and eventually expelled is 
four times the oxygen mass present. This is qualitatively 
in accordance with stellar evolution models which 
predict a more effective third dredge-up in low-$Z$  
stars (\eg, Ma\-rigo et al. \cite{margir99}), as observationally 
confirmed in the Magellanic Clouds (\eg, Leisy \& Dennefeld
\cite{leiden96}). Adding up C+N+O (sum left unchanged by 
CNO bi-cycle) for Wray~16-423 and He~2-436 shows that 
the latter had 40\% more dredge-up per hydrogen atom. 
The amount of third dredge-up material may depend on the 
timing of the last thermal pulse(s). For low-mass stars, 
a few dredge-up episodes are responsible for most of the enrichment 
(\eg, Wood \cite{wood97}). 

The nearly solar value of N/O in He~2-436 is in rough agreement with 
expectation as the initial N/O might have been of order 
one third the solar value in this moderate-$Z$ galaxy 
and the predicted first-dredge-up enrichment of nitrogen is then 
by a factor 2-3 after ``Cool Bottom Processing'' 
(CBP, Boothroyd et al. \cite{boosac99}). 
The nitrogen `enrichment' of Wray~16-423 (compared to He~2-436) points to a  
difference between the two PNe. Note that, with He/H = 0.108 and 
N/O = 0.22, Wray~16-428 is close to the edge of the PN Type~I class, 
defined in the LMC as He/H $>$ 0.105 and N/O $>$ 0.3 
(Torres-Peimbert \& Peimbert \cite{torpei97}). 

This nitrogen enrichment is on the order of the possible oxygen depletion 
(0.03dex) if both PNe originally had the same abundances. 
At temperatures above 2.5$\times$10$^{7}$K, 
while the ON cycle reaches equilibrium, O is transformed into N. 
However, since the CN cycle first reaches equilibrium, nitrogen should 
already be much enriched before this can hold 
(\eg, Smith et al. \cite{smishe97}). Alternatively, similar, 
relatively large nitrogen enrichment may have occured in both stars, 
nitrogen being then selectively converted into heavier elements,  
following high-temperature processing in the He~2-436 precursor 
under He-burning conditions (carbon is more abundant in He~2-436). 
However this interpretation is hampered by the fact the 
neon abundance appears to be almost the same in both PNe. 

Finally, although nitrogen enrichment due to CNO bur\-ning of freshly 
synthesized carbon at the base of the 
convective envelope is normally restricted to massive AGB stars (``Hot 
Bottom Burning'', \eg, Bl\"{o}cker \& Sch\"{o}nberner \cite{blosch91}), 
the excess nitrogen in Wray~16-423 is a minute fraction of the 
carbon enrichment so that it possibly results from a 
transformation of carbon before third dredge-up. 

In Table~7, average logarithmic abundance 
depletions $<$O, Ne$>$, $<$O, Ne, S$>$, $<$O, Ne, S, Ar$>$  and 
$<$S, Ar$>$ with respect to solar are given for the Sagittarius PNe. 
In Wray~16-423, all of these averages are obviously 
consistent with each other. The ``best'' averages, $-0.575\pm0.03$ 
and $-0.515\pm0.04$ for the old and new solar abundance sets respectively, 
are also consistent with the averages for all 7 elements 
($-0.54\pm0.08$ and $-0.50\pm0.07$ respectively). 

On the other hand, there seems to be a trend for increasing depletion with 
atomic number in He~2-436. The difference 
$\delta=$~$<$O,~Ne$>$~$-$~$<$S,~Ar$>$ is marginally significant for 
both solar sets ($\delta=0.16\pm0.11$ and $\delta=0.13\pm0.07$ resp.), 
although the interpretation may depend on the set used.  
Adopting the 1989 set, the depletion of sulfur is more pronounced 
but still compatible with those of oxygen and neon, 
with the stronger depletion of argon left as 
The argon abundance adopted in the He~2-436 models 
(Tables~3 and 5) is ``conservatively large'' since \ariii\ is correctly 
fitted and \ariv\ overestimated (Table~4), then suggesting a genuine 
underabundance of argon, at least compared to O and Ne. By contrast, 
adopting the 1998 set, 
an enrichment of O and Ne with respect to both S and Ar would be suggested. 

Other combinations of depletions 
are possible however since there is no relationship between the 
decrease of, \eg, the solar O and Ar abundance determinations 
from 1989 to 1998. Within quoted uncertainties of the 1998 solar 
abundances ($\pm0.06$), the value of $\delta$ for He~2-436 can 
be reduced to less than 0.10 and yet our previous conclusion that 
depletions in Wray~16-423 are identical for all elements beyond 
nitrogen still apply. Since a $\delta < 0.10$ is manageable 
within the combined uncertainties arising from models (Table~6), 
it would be premature to conclude 
that S and/or Ar are ``depleted'' with respect to O and Ne in 
He~2-436. The suggestion concerning {\sl solar} abundances is that the 
most consistent set of depletions is obtained in the Sagittarius PNe 
if O and Ar are intermediate between the 1989 and 1998 
values and S close to the lower end of the range quoted in Table~6. 
Then all of the abundances available for oxygen and heavier elements 
in both PNe are consistent with a depletion of $-0.55\pm0.07$. 
More specific conclusions require more definite reference abundances. 

\begin{table}
\caption{Average depletions with respect to solar abundances}
\begin{flushleft}
\begin{tabular}{lcccc}
\hline 
     & \multicolumn{2}{c}{Wray~16-423 -- $\odot$ } & \multicolumn{2}{c}{He~2-436 -- $\odot$ } \\
Elements    &   $\odot$ 89  & $\odot$ 98    &     $\odot$ 89 & $\odot$ 98 \\
\hline
O, Ne       & --.57$\pm$.04 & --.52$\pm$.02 &  --.56$\pm$.02 & --.51$\pm$.05 \\
O, Ne, S    & --.57$\pm$.03 & --.53$\pm$.04 &  --.59$\pm$.05 & --.55$\pm$.09 \\
O, Ne, S, Ar& --.58$\pm$.03 & --.51$\pm$.05 &  --.64$\pm$.10 & --.57$\pm$.08 \\
S, Ar       & --.59$\pm$.03 & --.51$\pm$.09 &  --.72$\pm$.09 & --.64$\pm$.02 \\
\hline
\end{tabular} 
\end{flushleft}
\end{table}

This re-analysis emphasizes the striking similarity of most abundances 
(excepting C and N) in both PNe. 
This ``chemical homogeneity'' is to be contrasted with the large 
abundance spread (0.7dex) implied by stellar studies (Mateo \cite{mat98}). 
Also striking is the contrast with the wide range of abundances among 
Galactic Halo PNe (\eg, Howard et al. \cite{howhen97}). Compared to 
Galactic PNe (Kingsburgh \& Barlow \cite{kinbar94}), a depletion 
$\sim$~0.4dex is obtained, but with large scatter 
(0.28 and 0.53dex for S and Ne respectively). 
However, considering the large discrepancy 
between the empirical and model S abundances in the case of the 
Sagittarius PNe (Table~6), doubts may be expressed about the 
reality of the low sulfur abundance of Galactic PNe, also 
based on empirical methods. The scatter would be reduced if this 
abundance was underestimated in the Galactic sample. 

Despite the similarities shown by the two Sagittarius PNe, 
some differences are apparent. 
There is a {\sl tendency} for oxygen to be more abundant in He~2-436 
than in Wray~16-423, even though the error bars  
overlap (P\'equi\-gnot et al. \cite{peqzij00}). 
Also, the argon abundance looks 
unexpectedly smaller in He~2-436 than in Wray~16-423 (factor 1.5)
and it is not clear how the model imperfections may gravely invalidate 
the significance of a direct comparison. 

\subsection{The nature of the nucleus of Wray~16-423}

In Paper~I, Wray~16-423 was classified as type [WC8] on the basis of the 
similar strengths of the broad [WR] features \ciiip\ 465.0\nm\ and 
\civ\ 580.6\nm. This classification was puzzling, considering the 
high Zanstra temperature of the star, 
more than confirmed by the present models. 

According to Paper~I (Table~9, fluxes not de-reddened), the total 
flux of the $\lambda$464.5 blend in Wray~16-423 is 1.78 (\Hb\ = 100). 
Assuming a \civ\ intensity ratio \la465.8/\la580.6 as in He~2-436, the 
contribution of the [WR] feature \civ\ 465.8\nm\ to the blend should be 0.21. 
Nebular line contributions are as follows. 
The Bowen fluorescence multiplet \niii\ 463.4-4.2\nm\ is $\sim$ 0.91 
(8\% of \heii\ 468.6\nm, comparing to typical high-excitation PNe). 
The recombination multiplet \ciiip\ 464.7-5.2\nm\ is about twice 
as strong as \ciiip\ 418.7\nm\ (Table~2), or 0.27. The recombination 
multiplet \oiip\ 464.0-466.0\nm\ is 0.24 (Table~2). The grand total 
is 1.63, which compares quite well with the observed flux. 

Thus, within uncertainties, there is no need for a [WR] \ciiip\ 
contribution to the $\lambda$464.5\nm\ feature and Wray~16-423 can now 
be an early-type [WC], as expected. Higher resolution observations may 
re-inforce this contention. 

\subsection{Stars in the HR diagram}

\begin{table}
\caption{Central star properties}
\begin{flushleft}
\begin{tabular}{lcccc}
\hline 
 &    \multicolumn{2}{c}{Wray~16-423}  &   \multicolumn{2}{c}{He~2-436} \\
 &W97$^a$ & \multicolumn{1}{c}{Models} &W97$^a$ &\multicolumn{1}{c}{Models} \\
\hline
\Teff/10$^3$K & 85 -- 96 & 107$\pm$10 & 61 & 70$\pm$10 \\
\Lstar/10$^3$\Lsun &  2.3  & 4.35$\pm$0.30$^b$ & 5.3  &  5.40$\pm$0.75$^b$ \\
\hline
\end{tabular} 

$^a$ Walsh et al. (\cite{waldud97}), Paper~I\\
$^b$ The $\pm$2~kpc uncertainty on the distance is not included.\\
\end{flushleft}
\end{table}

The central star properties turned out to be very similar 
in the HC and NEBU models (Sects.~3.3.3 ans 4.3.3) for both PNe. 
Effective temperatures \Teff\ and luminosities \Lstar\ 
are listed in Table~8 and compared 
to the empirical results of Paper~I. The error bars are the 
sums of the uncertainties on modeling and absolute 
flux measurement, excluding (systematic) 
errors on the distance to the galaxy. 
\Lstar(Wray~16-423) is probably very accurate, owing to the perfect 
agreement of either the different absolute fluxes or the different 
model results. The new determination is significantly larger than 
the previous one, which had to rely on difficult continuum measurements 
and did not take into account the leakage of ionizing radiation. 
The uncertainty is larger for \Lstar(He~2-436), considering that 
(1) the radio flux predictions are sensitive to details of 
geometry (self-absorption), (2) helium is a problem in the HC model 
and (3) the conversion of $L_B$ (NEBU model) into \Lstar\ is not 
straightforward. The agreement with Paper~I, 
based on the uncertain \Hb\ flux, is partly fortuitous.

The determination of \Lstar\ and \Teff\ allows the central stars to be 
placed on an HR diagram (Fig.~1). The theoretical evolutionary tracks 
plotted in Fig.~1 are taken from Vassiliadis and Wood (\cite{vaswoo94}) 
using $Z$ = 0.004, in accordance with the elemental abundances of the PNe. 
It was suggested in Paper~I that the 
central stars were on He-burning tracks, as they showed [WC] spectral 
signatures and differed by a factor 2 in luminosity. 
The radio continuum data and the photoionization models showed that 
both luminosities \Lstar\ were in fact rather large and differed 
by a factor of only 1.24$\pm$0.24, the smaller 
\Lstar\ of Wray~16-423 being explainable by its proximity to 
the turn-over. The positions of the stars on the HR diagram are now 
compatible with them being on H-burning tracks of 
initial masses $M$ = 1.19$\pm$0.08 and 1.22$\pm$0.14~\Msun\ 
for Wray~16-423 and He~2-436 respectively. Both stars may belong to 
the same evolutionary track. Then the weighted stellar-remnant mass is 
0.61~\Msun\ and the initial mass: 

$M$ = (1.19$\pm$0.10)~\Msun. 

\begin{figure}[ht]
\resizebox{\hsize}{!}{\includegraphics{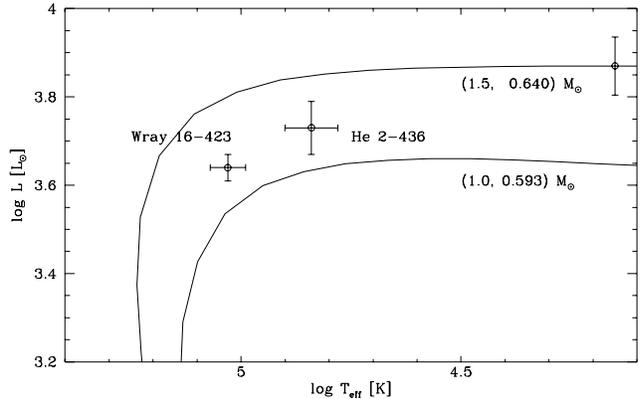}}
\caption[]{HR diagram for Wray~16-423 and He~2-436 and evolutionary tracks for H-burning stars with $Z$ = 0.004 according to Vassiliadis and Wood (\cite{vaswoo94}). 
The error bar attached to the 1.5~\Msun\ track corresponds to the $\pm$2~kpc uncertainty on the distance to Sagittarius. 
}
\end{figure}

For stars belonging to 
``standard'' He-burning tracks of Vassiliadis and Wood (\cite{vaswoo94}), 
the initial masses would be about 1.7~\Msun, too large 
to be reconciled with AGB star properties in Sagittarius (Sect.~4.8). 
Evolutionary time scales are such however that definite tracks 
specific to helium burning may not be relevant  
(Boothroyd \& Sackmann \cite{boosac88}; Bl\"ocker \cite{blocke95}), 
whereas statistical analyses support the notion that H-burning-track 
luminosities may be more appropriate even for [WR] stars, 
at least in the Galaxy (\eg, Gorny \cite{gorny00}). Since the 
Sagittarius PN nuclei are probably He-burners, the genuine 
initial mass of both stars may be somewhat larger than in 
the above estimate. 

\subsection{Evolutionary status}

The kinematic time scale of a PN may not faithfully reflect its 
``age'', due to uncertainties on the pre-PN wind velocity and 
the different acceleration processes. 
Comparing directly the two PNe may provide a more reliable kinematic 
time. The average hydrogen density of Wray~16-423 is $<$\Nh$>$ 
$= 0.44\times10^4$cm$^{-3}$. In He~2-436, assuming spherical symmetry 
for the outer shell (of which only 1/3 is currently illuminated 
due to shielding) and adding the mass of the inner shell, 
then $<$\Nh$>$ $= 2.7\times10^4$cm$^{-3}$. The weighted average 
expansion velocities of He~2-436 and Wray~16-423 are 
14.1\kms and 25.5\kms respectively (Gesicki \& Zijlstra \cite{geszij00}). 
Assuming volume expansion with mean velocity (14.1+25.5)/2~\kms 
from the present outer radius $\sim$~1.1$\times10^{17}$cm of He~2-436, 
the density ratio of the PNe translates into a time scale $\Delta t_{kin}$ 
= 1430~yr, with ``final'' radius $\sim$~2.0$\times10^{17}$cm. The actual 
average radius of Wray~16-423 is larger ($\sim$~$2.4\times10^{17}$cm) 
as more mass can be ionized owing to expansion. Meanwhile the dense 
inner shell of He~2-436, whose current kinematic time scale is probably 
less than 1000~yr, will fade considerably, as its density will decrease 
by a factor $\sim$~20, making He~2-436 look like Wray~16-423 at present. 

The dynamic time scales of the stars are uncertain too. Given 
the core mass, the early post-AGB evolution is primarily sensitive to 
the poorly known mass loss rate. Although the evolutionary tracks of 
Vassiliadis \& Wood (\cite{vaswoo94}) are relevant to determine core 
and main-sequence masses (Fig.~1), their post-AGB wind strength may be 
underestimated (Wood \cite{wood97}; Sch\"onberner \cite{schoen97}). 
On the other hand, Bl\"ocker (\cite{blocke95}) probably overestimated 
mass loss rates during the early AGB evolution but tried to describe 
more realistically the decrease of mass loss as \Teff\ increases 
during post-AGB and his results are preferred for time scales. 
For core mass 0.61~\Msun\ and the large \Teff's relevant to the 
present PN nuclei, standard parametric mass losses apply, 
leading to d\Teff/d$t$~$\sim$~25K/yr for H-burning and a time scale 
$\Delta t_{dyn}$ = (1500$\pm$800)~yr (the uncertainty arising mainly from 
the \Teff\ determinations), in accord with $\Delta t_{kin}$. 
Although the evolution of helium burners is sensitive to the epoch at which
the fatal thermal pulse ignites, and there are no model tracks 
for such an evolution, the regime of \Teff's of interest here 
(in between the post-AGB and cooling phases) may be 
propitious to time scales relatively insensitive to 
details of the past history of the stars. 
Despite very large uncertainties, this coincidence of 
time scales re-inforces the analogy of these PNe. 

The nucleus of He~2-436 is a [WC4] star (Paper~I). This early type 
suggests an advanced stage of evolution, at variance with the {\sl apparent} 
youth of the PN: the \oiii-weighted mean electron density \Ne\ 
is over 10$^5$~cm$^{-3}$, met among Galactic PNe with 
[WR] central stars only for the latest possible type [WC11] 
(\eg, Gorny \cite{gorny00}). 
Photoionization models show that the average \Ne\ of He~2-436 is in fact 
a few 10$^4$~cm$^{-3}$ (Sect.~4.7), alleviating the difficulty 
(Galactic [WC4] stars have \Ne\ up to 10$^4$~cm$^{-3}$). 
Wray~16-423, also excited by an early-type [WC] star, 
corresponds to the high-\Ne\ end of Galactic counterparts. 

\subsection{Stellar population}

The youngest stars in Sagittarius have ages around 5 Gyrs 
(Mateo \cite{mat98}) which would give a mass of around 1.3~\Msun, 
in fair agreement with the values above (Sect.~4.6). 
This coincidence of initial masses applies insofar as both PN nuclei have 
luminosities typical of H-burning or slightly lower. 
Star formation in dwarf galaxies is thought to occur
in bursts (\eg, Smecker-Hane et al. \cite{smeck}), but 
the level of star formation in between is not well known.
In view of the strong similarities of the abundances,
little evolution of the ISM can have taken place between the formation
of the two stars, which favours an origin within the same star
formation burst in a well-mixed ISM. 

The carbon stars in Sagittarius have well-determined magnitudes peaking 
at $M_{bol}=-4.5$ (Whitelock et al. \cite{whit96}), corresponding to 
$L=5 \times10^3$~\Lsun, in 
extremely good agreement with the two PNe. Together
with the high C/O of the PNe, this is strong evidence that the carbon
stars form the source population of the PNe. 

The abundances of the two PNe are independent of the age-metallicity
degeneracy which affects photometric abundance determinations of the
stellar population. They are therefore an excellent way of calibrating
the photometric results.
>From the AGB, Whitelock et al. (\cite{whit96}) estimated 
[Fe/H]=$-0.8$. Ibata et al. (\cite{ibagil94}) 
found a mean [Fe/H] of $-1.1$. Marconi et al. (\cite{marbuo98}) 
derived a metallicity range between $-0.71$
and $-1.58$. Sarajedini \& Layden (\cite{sarlay95}) found several
populations, the most metal-rich of which having [Fe/H] = $-0.52$.
Only this last value corresponds to the PNe ($-0.55\pm0.07$). 

It should be noted that, from the slope of the giant branch, 
Whitelock et al. (\cite{whit96}) derived a larger abundance ($-0.58$) 
than the one they finally adopted. 
The difference was related to an assumption of a population age 
around $10^{10}$ yr.  If a younger age is adopted, the AGB 
and PN abundances agree extremely well. It is therefore suggested 
that [Fe/H] of the AGB stars is in fact $-0.58$ rather than 
$-0.8$. The differences with other studies may be due to the 
age--metallicity degeneracy, 
the most metal-rich population of 
Sagittarius being somewhat younger than assumed in these studies. 
The presence of metal-poor populations is not in doubt, 
but the result of Sarajedini \& Layden (\cite{sarlay95}) 
is confirmed regarding the most metal-rich population. 

\section{Conclusions}

Abundances for the two PNe in the Sagittarius 
dwarf galaxy have been improved. 
The abundances of both PNe now appear remarkably similar 
for elements heavier than nitrogen. Abundances of nitrogen and carbon 
demonstrate the operation of the first and third dredge-ups. The larger 
overabundance of carbon in \he\ is not unexpected as the total amount 
of third-dredge-up material may depend on the timing of 
the last thermal pulse(s). The excess nitrogen in 
\wray\ is more intriguing, considering that second dredge-up does not occur 
in low-mass stars and helium is equally abundant in both objects. 
The large helium overabundance may be another problem. 

The photoionization codes provide well-defined 
results for Wray~16-423 and a larger range of acceptable 
solutions for He~2-436. In the latter case, the unicity of solution 
breaks down, as a consequence of the high density in the 
inner shell. Line fluxes are matched within the 
1-$\sigma$ statistical errors, that is 
only a few percent for most basic lines, whilst the degrees of freedom 
are outnumbered by the independent observables.

Despite their [WC] spectra, both central stars delineate the 
1.2~\Msun\ H-burning track, in fair agreement with the mass of 
the carbon stars of Sagittarius. The abundances of the PNe may 
help to calibrate the stellar photometry 
of the more metal-rich population in this galaxy. 

Improving the absolute \Hb\ and radio fluxes for \he\ and, particularly, 
observing at radio frequencies above 9~GHz would more tightly constrain 
the geometry of the nebula and the luminosity of the nucleus, then 
leading to a more accurate mass of the star. 
Concerning \wray, the critical parameter to improve the 
positioning on the HR diagram is the effective temperature of the nucleus. 
Model atmospheres describing the ionizing FUV radiation from hot 
[WC] stars are needed. 

The small angular size of the Sagittarius PNe allowed to secure global, 
reproducible spectra. The spectra obtained by Walsh et al. (\cite{waldud97}) 
proved to be reliable in most of the optical range, as demonstrated 
by the overall excellent agreement between theory and 
observation for the \hi\ and \hei\ spectra (Table~6). They can 
however be usefully improved in the blue and near-UV and completed in 
the far-red and the UV. UV spectra are critical to check the energy 
balance and the carbon abundance of both PNe. 
New deeper optical observations are possible and highly desirable, 
considering that these objects 
represent a moderate-$Z$ stellar population and appear as 
excellent test beds of models. Predictions for 
recombination line fluxes must be checked to ensure that these PNe 
are not subject to the same kind of discrepancy noted in some 
Galactic PNe between abundances obtained from collisional and 
recombination lines (\eg, Liu et al. \cite{liusto95}).

\end{document}